\documentclass[a4paper,11pt]{article}
\usepackage{jcappub}
\usepackage{lineno}
\usepackage{amsmath,amssymb}
\usepackage[dvipsnames]{xcolor}
\hypersetup{
    colorlinks=true,
    linkcolor=Maroon,
    citecolor=MidnightBlue,
    urlcolor=blue
}
\usepackage{graphicx}
\usepackage{slashed}
\usepackage{bm}
\usepackage{mathrsfs}
\usepackage{multirow}
\usepackage{makecell}
\newcommand{\dE}[3][(nlm)]{\dot{#2}_\mathrm{#3}^{#1}}
\newcommand{\M}[3][(nlm)]{{#2}_\mathrm{#3}^{#1}}
\newcommand{\nbar}{\bar{n}}
\newcommand{\rcite}[1]{Ref.~\cite{#1}}
\newcommand{\rscite}[1]{Refs.~\cite{#1}}
\newcommand{\eref}[1]{Eq.~\eqref{#1}}
\newcommand{\fref}[1]{Fig.~\ref{#1}}
\newcommand{\beq}{\begin{equation}}
\newcommand{\eeq}{\end{equation}}
\newcommand{\bea}{\begin{eqnarray}}
\newcommand{\eea}{\end{eqnarray}}

\allowdisplaybreaks
\arxivnumber{2501.09280}
\title{Effect of accretion on scalar superradiant instability}
\author[a]{Yin-Da Guo,}
\author[a]{Shou-Shan Bao,}
\author[b,c,d]{Tianjun Li}
\author[a]{and Hong Zhang}
\affiliation[a]{Institute of Frontier and Interdisciplinary Science,
Key Laboratory of Particle Physics and Particle Irradiation (MOE),
Shandong University, Qingdao 266237, China}
\affiliation[b]{CAS Key Laboratory of Theoretical Physics, Institute of Theoretical Physics,
Chinese Academy of Sciences, Beijing 100190, China}
\affiliation[c]{School of Physical Sciences, University of Chinese Academy of Sciences, No. 19A Yuquan Road, Beijing 100049, China}
\affiliation[d]{School of Physics, Henan Normal University, Xinxiang 453007, China}

\emailAdd{yinda.guo@mail.sdu.edu.cn}
\emailAdd{ssbao@sdu.edu.cn}
\emailAdd{tli@itp.ac.cn}
\emailAdd{hong.zhang@sdu.edu.cn}
\abstract{
Superradiance can lead to the formation of a black hole (BH) condensate system. We thoroughly investigate the accretion effect on the evolution of this system, and the gravitational wave signals it emits in the presence of multiple superradiance modes. Assuming the multiplication of the BH mass and scalar mass as a small number, we obtain the analytical approximations of all important quantities, which can be directly applied to phenomenological studies. In addition, we confirm that accretion could significantly enhance the gravitational wave (GW) emission and reduce its duration, and show that the GW beat signature is similarly modified.
}
\begin{document}
\maketitle
\flushbottom
%
%%%%%%%%%%%%%%%%%%%%%%%%%%%%%%%%%%%%%%
\section{Introduction}
Superradiance allows ultra-light bosons to extract energy and angular momentum from rotating black holes (BHs). This mechanism exists when the condition $\omega < m\Omega_\mathrm{H}$ is satisfied, where $\omega$ is the frequency of the boson field, $m$ is the magnetic number, and $\Omega_\mathrm{H}$ is the angular velocity of the event horizon. This process leads to a significant accumulation of bosons in bound states surrounding a Kerr BH, forming a BH-condensate system analogous to a hydrogen atom. For a comprehensive review on superradiance, we refer the readers to \rcite{Brito:2015oca}. In this work, we study the superradiance of ultra-light scalars, such as the QCD axion \cite{Weinberg:1977ma,Wilczek:1977pj}, {\color{black} and axion-like particles suggested by string theory \cite{Arvanitaki:2009fg}}. 

Many observables are proposed to observe such BH-condensate systems \cite{Chen:2019fsq,Chen:2021lvo,Blas:2020nbs}. In this article, we focus on the gravitational effects of the scalars, independent of how the scalar field couples to the Standard Model particles. Both direct and indirect detection methods have been proposed in the literature. One could directly observe the gravitational wave (GW) signals from the BH-condensate system, including quasi-monochromatic GWs \cite{Arvanitaki:2010sy,Yoshino:2013ofa,Yoshino:2014wwa,Arvanitaki:2014wva,Arvanitaki:2016qwi,Brito:2017wnc,Brito:2017zvb,Baryakhtar:2017ngi,Guo:2022mpr}, stochastic GWs \cite{Brito:2017wnc,Brito:2017zvb} and GW beats \cite{Guo:2022mpr}. The consequence of the BH superradiance could also be indirectly studied via the mass-spin distribution of numerous BHs, which would align along Regge trajectories if superradiance occurs \cite{Arvanitaki:2010sy,Cardoso:2018tly,Fernandez:2019qbj,Ng:2019jsx,Ng:2020ruv,Cheng:2022jsw}.

Both indirect and direct searches require a thorough understanding of the evolution process of the BH-condensate system. In previous studies, accretion is often overlooked for simplicity. The obtained results are then qualified only for isolated BHs, without gas and stars in the neighborhood. Nonetheless, accretion is important in the evolution of most BHs, even crucial for those BHs in active galactic nuclei or X-ray binaries. To our best knowledge, the impact of gas accretion on the superradiant evolution was first pointed out in \rcite{Brito:2014wla} and further studied in Refs.~\cite{Fukuda:2019ewf,Roy:2021uye,Hui:2022sri,Unal:2023yxt,Sarmah:2024nst}.

In this work, we carefully investigate the effect of accretion on the evolution of BH-condensate systems. We numerically solve the evolution equations with both the dominant and the subdominant modes. Analytical approximations of important quantities and time scales are also provided and compared with the numerical results. We then discuss the effects of accretion on the GW signals, including GW beat, which was proposed to distinguish the GW from superradiance and other monochromatic sources~\cite{Guo:2022mpr}. These numerical and analytical results provide baselines for more complicated cases, such as those with nonzero couplings between the ultralight scalar and the SM particles.

This paper is organized as follows. Sec.~\ref{sec:setup} begins with a brief review of scalar superradiance, followed by discussions on the evolution of BHs without the scalar field or accretion. Then in Sec.~\ref{sec:single}, we study the effect of accretion on evolution with solely the dominant mode and provide analytic estimates of the essential quantities. This approach is then generalized to the scenario with multiple modes in Sec.~\ref{sec:multi}, with a brief discussion of the impact of accretion on GW signals. Finally, we summarize our findings in Sec.~\ref{sec:summary}.

Throughout the paper, we adopt the Planck units in which $G=\hbar=c=1$, and use the $\{+,-,-,-\}$ signature.

%%%%%%%%%%%%%%%%%%%%%%%%%%%
%%%%%%%%%%%%%%%%%%%%%%%%%%%
\section{Setup}\label{sec:setup}
\subsection{Scalar superradiance}

The Kerr spacetime metric with mass $M$ and angular momentum $J$ is expressed in the Boyer-Lindquist coordinates as \cite{Boyer:1966qh}
%%%%%=====
\begin{equation}
    \begin{aligned}
        ds^2=&\left( 1-\frac{2Mr}{\Sigma} \right) dt^2+\frac{4aMr}{\Sigma}\sin ^2\theta dtd\varphi -\frac{\Sigma}{\Delta}dr^2
        \\
        &-\Sigma d\theta ^2-\left[ \left( r^2+a^2 \right) \sin ^2\theta +2\frac{Mr}{\Sigma}a^2\sin ^4\theta \right] d\varphi ^2,
    \label{eq:KerrMetric}
    \end{aligned}
\end{equation}
%%%%%
where
%%%%%=====
\begin{subequations}
\begin{align}
    a & \equiv J/M,\\
    \Delta & \equiv r^2-2 M r+a^2,\\
    \Sigma & \equiv r^2+a^2 \cos ^2 \theta.
\end{align}
\end{subequations}
%%%%%
The inner horizon $r_-$ and the outer horizon $r_+$ are respectively defined as
%%%%%=====
\begin{equation}
    r_{ \pm}=M \pm \sqrt{M^2-a^2}.
\end{equation}
%%%%%
The variable $a$ is the angular momentum of unit BH mass. Throughout this paper, we use the dimensionless variable $a_* \equiv a/M$ and refer to it as the BH {\it spin} for brevity.

In this work, we consider a real scalar field surrounding a Kerr BH. Due to the small energy density of the condensate, its backreaction to the metric and self-interaction of the scalar field can be safely ignored \cite{Brito:2014wla}. They can be included systematically as perturbations if high precision is required, which is beyond the scope of this work. Below we study a free real scalar field on the Kerr metric
%%%%%=====
\begin{equation}
	\left(\nabla^{\rho}\nabla_{\rho}+\mu^{2}\right)\Phi=0,
	\label{eq:_KG_equation}
\end{equation}
%%%%%
where $\mu$ is the scalar mass. Generally, the energy eigenvalues are complex numbers depending on three indices, which are the overtone number $n$, azimuthal number $l$, and magnetic number $m$. The principal number $\bar{n}=n+l+1$ is also widely used instead of $n$. For eigenstate with indices $\{n,l,m\}$, we use $\omega_{nlm}$ and $\Gamma_{nlm}$ to denote the real and imaginary parts of the eigenvalue, respectively. For bounded scalar fields, the $\omega_{nlm}$ can be expressed as a series of $\alpha=M\mu$~\cite{Baumann:2019eav}
%%%%%=====
\begin{align}\label{eq:omega}
	\omega_{nlm} &= \mu \Big(1-\frac{\alpha^2}{2\bar{n}^2}-\frac{\alpha^4}{8\bar{n}^2}+\frac{f_{\bar{n}l}}{\bar{n}^3}\alpha^4+\frac{h_l a_* m}{\bar{n}^3}\alpha^5+\cdots\Big),
\end{align}
%%%%%
with
%%%%%=====
\begin{align}
	f_{\bar{n}l} & \equiv -\frac{6}{2l+1}+\frac{2}{\bar{n}},\\
	h_{l} & \equiv \frac{16}{2l(2l+1)(2l+2)}.
\end{align}
%%%%%
The imaginary part $\Gamma_{nlm}$, which is also called the superradiance rate, has been calculated at the leading order in Ref.~\cite{Detweiler:1980uk} and has been recently improved to next-to-leading order (NLO) in Ref.~\cite{Bao:2022hew}. Both the expressions have compact forms while the latter is much more accurate compared with the numerical calculation in Ref.~\cite{Dolan:2007mj}. In this work, we use the NLO expression, which gives\footnote{This expression is rewritten from the result in Ref.~\cite{Bao:2022hew}.}
%%%%%=====
\begin{align}\label{eq:Gamma}
	\begin{split}
		& \Gamma_{nlm} =
		-\omega_1 \left(4\kappa\sqrt{M^2-a^2}\right)^{2l^\prime+1}\frac{\Gamma (n + 2 l^\prime + 2)}{n!}\frac{\sinh (2 \pi  p)}{2 \pi }
		\\
		& \hspace{1.2cm} \times\frac{\left| \Gamma \left(l^\prime+1-i p+\sqrt{q-p^2}\right) \Gamma \left(l^\prime+1+i p+\sqrt{q-p^2}\right) \right| ^2}{\left[\Gamma ( 2 l^\prime + 1)\Gamma ( 2 l^\prime + 2)\right]^2},
	\end{split}
\end{align}
%%%%%
where $l^\prime \equiv l+\epsilon$, $p \equiv M r_+ (\omega_{nlm} - m \Omega_\text{H}) / \sqrt{M^2-a^2}$, $\kappa \equiv \sqrt{\mu^2-\omega_{0}^2}$, $ \Omega_\mathrm{H} \equiv a/(2Mr_+)$, and
%%%%%=====
\begin{subequations}
\begin{align}
	\epsilon & \equiv -\frac{8\alpha^2}{2l+1},
	\\
	q & \equiv \frac{8 M r_+ \omega_{0} (r_+\omega_{0} - m M \Omega_\text{H})}{r_+ - r_-} - \mu^2 (r_+^2 + a^2) +4 M^2 (\mu^2 - 3 \omega^2_{0}),
	\\
	\omega_0 & \equiv \mu \sqrt{1 - \frac{2 \alpha^2}{\nbar^2 +4\alpha^2 + \nbar \sqrt{\nbar^2 + 8 \alpha^2}}},
	\\
	\omega_1 & \equiv \frac{\mu^2 - \omega_{0}^2}{\nbar \omega_{0} (1 + 4 M^2 (\omega_{0}^2 - \mu^2) / \nbar)}.
\end{align}
\end{subequations}
%%%%%

Eq.~\eqref{eq:Gamma} indicates that the superradiance condition $\Gamma_{nlm}>0$ is equivalent to $\omega < m \Omega_\text{H}$, from which one could derive the critical BH spin for each mode labeled by $\{n,l,m\}$
%%%%%=====
\begin{align}\label{eq:a_critc}
    a^{(nlm)}_\mathrm{*c}=\frac{4mM\omega_{nlm}}{m^2+\left( 2M\omega_{nlm} \right) ^2}\text{, for $M\omega_{nlm}\leq\frac{m}{2}$}.
\end{align}
%%%%%
The superradiance of $\{n,l,m\}$ mode occurs only when the BH spin $a_*$ is greater than $a^{(nlm)}_\mathrm{*c}$.

The superradiance rate could be very different for different $\{n,l,m\}$ modes.
In general, the most important modes are those with $l=m$ which is the smallest integer greater than $\omega_{nlm}/\Omega_H$. Among these modes, the superradiance rate is usually smaller with a larger value of $n$. This hierarchy is strict for $l=1$ and 2. For modes with $l\geq 3$, mode with larger $n$ may have a larger superradiance rate at some ranges of $M\mu$ and $a_*$, which is named ``overtone mixing" in the literature \cite{Siemonsen:2019ebd}. In this work, we first focus on the case in which modes with larger $n$ have smaller superradiance rates, obtaining analytical approximations of all important quantities. Then in the last section, we argue that these approximations also work in the presence of ``overtone mixing".

\subsection{Evolution equations} 

The evolution of the BH-condensate system includes the exchange of energy and angular momentum between the BH and the scalar field. The latter emits GW while rotating around the host BH. With accretion, the BH also absorbs energy from the accretion disk. In a more realistic scenario, the photons radiated by the accretion disk are important when the BH spin is very close to 1~\cite{Thorne:1974ve}. The evolution equations with all these effects are \cite{Brito:2014wla,Thorne:1974ve}
%%%%%=====
\begin{subequations} \label{eq:total_ODEs}
    \begin{align}
    	\dot{M}  &= \dot{M}_\mathrm{acc}\left(1  + \frac{E_\mathrm{rad}^{\dagger}}{E_\mathrm{ms}^{\dagger}}\right) - \sum_{nlm} \dE{E}{SR}, 
		\label{eq:M_BH_ODE} \\
    	\dot{J}  &= \frac{\dot{M}_\mathrm{acc}}{E_\mathrm{ms}^{\dagger}} \left(J_\mathrm{ms}^{\dagger} + J_\mathrm{rad}^{\dagger}\right) - \sum_{nlm} \frac{m}{\omega_{nlm}} \dE{E}{SR} , 
		\label{eq:J_BH_ODE} \\
    	\dE{M}{s} &=  \dE{E}{SR} -  \dot{E}_\mathrm{GW}^{(nlm)}, 
		\label{eq:Ms_ODE}\\
    	\dE{J}{s}  &= \frac{m}{\omega_{nlm}} \left(\dE{E}{SR} - \dE{E}{GW}\right),
        \label{eq:Js_ODE}
    \end{align}
\end{subequations}
%%%%%
where $M_s^{(nlm)}$, $J_s^{(nlm)}$, $\dE{E}{SR}$ and $\dE{E}{GW}$ are the mass, angular momentum, superradiance rate and the GW emission rate of the $\{n,l,m\}$ mode, respectively. The $\dE{E}{SR}$ is related to the imaginary part of the eigenfrequency by $\dE{E}{SR} \equiv 2 \M{M}{s} \Gamma_{nlm}$. The $\dE[]{M}{acc}$ is the mass absorption rate of the BH from the accretion disk. The $E_\mathrm{ms}^{\dagger}$ and $J_\mathrm{ms}^{\dagger}$ are the energy and angular momentum of unit mass in the last stable circular orbit, respectively. The accretion disk radiates photons, some of which are later captured by the BH. Its contribution to the BH mass and angular momentum are described by $E_\mathrm{rad}^{\dagger}$ and $J_\mathrm{rad}^{\dagger}$ for unit accreted mass, respectively, which are calculated in \rscite{Page:1974he,Thorne:1974ve}. The expressions for $E_\mathrm{ms}^{\dagger}$, $J_\mathrm{ms}^{\dagger}$ and $\dE{E}{GW}$ are listed in App.~\ref{sec:Approximate_expressions} for convenience.

In obtaining Eqs.~\eqref{eq:total_ODEs}, we have made the following assumptions:
%%%
\begin{itemize}
    \item 
    The backreactions of the condensate and the accretion disk on the metric, as well as the scalar self-interaction are ignored due to the low energy densities of the condensate and the disk \cite{Brito:2014wla}.
    \item 
    The quasi-adiabatic approximation is employed, which assumes that the dynamical timescale of the BH is much shorter than the timescales of both the accretion and superradiant instability~\cite{Brito:2014wla,Brito:2017zvb}.
    \item 
    The gravitational effect from other celestial bodies is neglected. The BH-condensate system with the accretion disk is effectively isolated. 
\end{itemize}
%%%
These contributions can be added perturbatively which are beyond the scope of this work. In the rest of this section, we study the evolutions without either the superradiance or accretion. Then we move on to the case with all these effects from the next section.

\subsubsection{The evolution without the scalar field}\label{sec:evo_wo_scalar}

In this part, we study the scenario in the absence of the scalar field. If the photon radiation can be further ignored, i.e. $E_\mathrm{rad}^{\dagger} = J_\mathrm{rad}^{\dagger}=0$, the evolution of the BH mass and angular momentum only depends on the accretion rate $\dot{M}_\text{acc}$. Useful results can be obtained even without any knowledge of $\dot{M}_\text{acc}$. From Eqs.~\eqref{eq:M_BH_ODE} and \eqref{eq:J_BH_ODE}, one could derive
%%%%%=====
\begin{align}\label{eq:das_dlnM}
	\frac{\mathrm{d}a_*}{\mathrm{d}\ln M} &= \frac{J_\mathrm{ms}^{\dagger}}{M E_\mathrm{ms}^{\dagger}}-2a_*.
	% \\
	% \frac{\mathrm{d}M}{\mathrm{d} M_\mathrm{acc}} &= E^{\dagger}_\mathrm{ms}.
\end{align}
%%%%%
BHs with constant spin $a_*=1$ and any value of $M$ is a trivial solution. For BH with initial spin $a_{*0}<1$ and mass $M_0$, a non-trivial solution also exist \cite{Thorne:1974ve}
%%%%%=====
\begin{align}
	a_*=
	\begin{cases}
		\sqrt{\frac{2}{3}}\frac{kM_0}{ M}\left(4-\sqrt{\frac{18k^2M_0^2}{M^2}-2}\right),&
		\text{if\ }  1 \leq \frac{M}{M_0} \leq {\sqrt{6}\,k},
		\\
		1,  &\text{if\ } \frac{M}{M_0} \geq {\sqrt{6}\,k},
	\end{cases}
	\label{eq:as_ini_not_0}
\end{align}
%%%%%
with
%%%%%=====
\begin{subequations}
	\begin{align}
	 k &= \frac{3 a_{*0} }{\sqrt{\frac{12 \sqrt{2}}{\sqrt{{3 \beta ^{4}}(1+a_{*0})^{-1}+3 \beta^{2}+2}}+18-\gamma ^2 }+2 \sqrt{6} -\gamma },\\
	 \gamma &= \beta^{-1}\sqrt{9(1-a_{*0}^2+\beta^{4})+6\beta^{2}},\\
	 \beta &= \sqrt[6]{(1+a_{*0})^2 (1-a_{*0})}.
	\end{align}
\end{subequations}
%%%%%=====
For the special case with $a_{*0}=0$, one obtains $k=1$ and the BH spin reaches its maximum value at $M=\sqrt{6}M_0$. For other values of $a_{*0}$, there is $0<k<1$. The BH with $a_{*0}<1$ evolves along the nontrivial solution, with both mass and spin increasing with time, until the spin reaches 1. Then the spin is fixed at 1 with mass increasing monotonically.

To derive a specific form of $\dot{M}_\text{acc}$, we first study the accretion with the maximum luminosity, i.e. the Eddington luminosity $L_\mathrm{Edd}$. Ignoring the BH capture of photons radiated from the accretion disk, the BH mass evolves as
%%%%%=====
\begin{align}\label{eq:Ledd}
	\frac{\dot{M}_\mathrm{acc}\eta}{1-\eta} = L_\mathrm{Edd} \equiv \frac{4\pi M\mu_\mathrm{p}}{\sigma_\mathrm{T}},
\end{align}
%%%%%
where $\eta$ is the efficiency of photon radiation from the accretion disk, $\sigma_\mathrm{T}$ is the Thompson cross-section, and $\mu_\mathrm{p}$ is the proton mass. For a thin accretion disk, the efficiency has a simple form $\eta=1-E_\mathrm{ms}^{\dagger}$~\cite{Thorne:1974ve}. For a Schwarzschild BH, $\eta$ is approximately $0.06$, while for an extreme Kerr BH with $a_*=1$, the value of $\eta$ can reach the value of $0.4$. Due to the small variation of $\eta$, it is usually assumed to be a constant for simplicity. According to Eq.~\eqref{eq:Ledd}, the BH mass then increases exponentially with a time scale
%%%%%=====
\begin{align}
	\tau_\mathrm{S} \equiv \frac{\sigma_\mathrm{T}}{4\pi \mu_\mathrm{p}} \cdot \frac{\eta}{1-\eta} \approx 4.5 \times 10^{8}\,\mathrm{yr} \cdot \frac{\eta}{1-\eta},
\end{align}
%%%%%
which {\color{black} is also known as the Salpeter timescale \cite{Salpeter:1964kb}.}
 
In this work, we choose $\eta=0.1$ and further assume the actual accretion rate is smaller than the extreme case by a factor of $f_\mathrm{Edd}$. Then the mass absorption rate is
%%%%%=====
\begin{align}\label{eq:M_acc}
	\dot{M}_\mathrm{acc} = f_\mathrm{Edd} \frac{1-\eta}{\eta} L_\mathrm{Edd} \equiv M(t)/\M[]{\tau}{acc},
\end{align}
%%%%%
where the accretion timescale $\M[]{\tau}{acc}$ is
%%%%%=====
\begin{align}\label{eq:tau_acc}
	\tau_\mathrm{acc} \equiv \tau_\mathrm{S}/f_\mathrm{Edd}= 5\times 10^7 \mathrm{ yr}/\M[]{f}{Edd}.
\end{align}
%%%%%
Inserting Eq.~\eqref{eq:M_acc} into Eq.~\eqref{eq:M_BH_ODE}, one obtains a BH mass increasing exponentially with time
%%%%%=====
\begin{align}\label{eq:BH_mass_efold}
	M(t) = M_0\exp(t/\M[]{\tau}{acc}).
\end{align}
%%%%%

%%%%%FFFFF
\begin{figure}
	\centering 
	\includegraphics[width=0.48\textwidth]{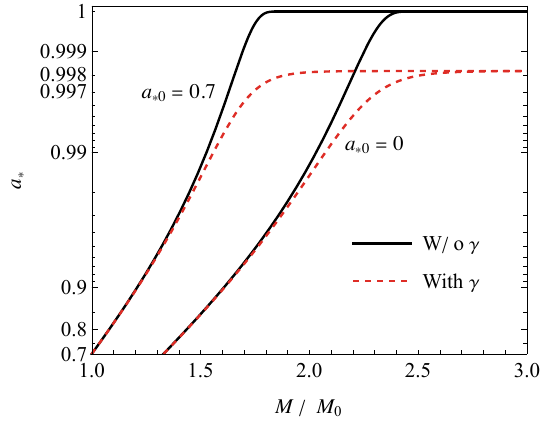}
	\caption{
The evolution of BH spin $a_*$ as a function of the mass ratio $M/M_0$ without the scalar field. Two initial BH spins $a_{*0}=0$ and $a_{*0}=0.7$ are considered. The red dashed curves are solved from Eq.~\eqref{eq:das_dlnM_with_photon}. The black curves further assumes the absence of the photon captured by the BH, which gives Eq.~\eqref{eq:as_ini_not_0}}. 
	\label{fig:das_dlnM} 
\end{figure}
%%%%%

Next, we add the contribution of photon capture by the BH. It has been shown that the photon-capturing rate of the BH is greater for photons with negative angular momentum, i.e. angular momentum opposite to the BH spin, than for photons with positive angular momentum \cite{Godfrey:1970am}. As a result, capturing photons emitted by the accretion disk slows down the increase of the BH spin. Mathematically, \eref{eq:das_dlnM} should be modified as follows \cite{Thorne:1974ve}:
%%%%%=====
\begin{align}\label{eq:das_dlnM_with_photon}
	\frac{\mathrm{d}a_*}{\mathrm{d}\ln M} &= \frac{1}{M} \frac{J_\mathrm{ms}^{\dagger} + \dE[]{J}{rad}}{E_\mathrm{ms}^{\dagger}+\dE[]{M}{rad}}-2a_*.
\end{align}
%%%%%
Using the expressions of $\dE[]{M}{rad}$ and $\dE[]{J}{rad}$ calculated in \rscite{Page:1974he,Thorne:1974ve}, we solve the equation numerically and present the solutions in Fig.~\ref{fig:das_dlnM}. The effect of photon-capturing is visible only when $a_* \gtrsim 0.97$. We also obtain a limiting value for the BH spin, $\M[]{a}{*lim} \approx 0.998189$, which is consistent with \rcite{Thorne:1974ve}. Since the photon capture's contribution to BH spin is smaller than $0.3\%$, its effect is always ignored in the analytic approximation throughout this work.

\subsubsection{The evolution without accretion}\label{sec:evo_wo_acc}

In the rest of this section, we study the scenario in the absence of accretion. Mathematically, the terms with $M_\text{acc}$ in Eqs.~\eqref{eq:total_ODEs} are taken to be zero. In this part, we consider a BH with initial mass $M_0$ and spin $a_{*0}$. The initial mass of the $\{n,l,m\}$ mode is set as $M_{\text{s},0}^{(nlm)}$. The $a_{*0}$ is assumed to be larger than $a_{*\text{c}}^{(111)}$ such that the dominant $\{0,1,1\}$ mode  and the subdominant $\{1,1,1\}$ mode grow due to superradiant instability. We focus on these two modes, with all other modes ignored. The evolution of this system can be separated into the spin-down phase, attractor phase of the subdominant mode, and attractor phase of the dominant mode. There has been extensive research on such isolated BH-condensate systems in the literature.  Below, we summarize the findings in Ref.~\cite{Guo:2022mpr}.

The spin-down phase starts at $t=0$ and ends at $t_1$, when the BH spin drops to $a_{*\text{c}}^{(111)}$. The BH mass decreases by a few percent during this time, with the mass deficiency transferred to the condensate. Both the $\{0,1,1\}$ and $\{1,1,1\}$ modes grow almost exponentially. At time $t_1$, the $\{1,1,1\}$ mode reaches its maximum mass. The masses of these two modes at $t_1$ could be estimated quite accurately with
%%%%%=====
\begin{align}
	\label{eq:Ms011_max_WoAcc}
	\frac{\M[(011)]{M}{s,1}-\M[(011)]{M}{s,0}}{M_0} & \approx M_0 \mu \left[ a_{*0} -4 M_0\mu +7.89 a_{*0} (M_0\mu)^2 
	+ (-16.55+6.45a_{*0}^2)(M_0\mu)^3\right],
	\\
	\frac{\M[(111)]{M}{s,1}}{M_0} & \approx \frac{\M[(111)]{M}{s,0}}{M_0}\left[\frac{\M[(011)]{M}{s,1}}{\M[(011)]{M}{s,0}}\right]^{0.35},
\end{align}
%%%%%
where $\M[(nlm)]{M}{s,1}\equiv\M[(nlm)]{M}{s}(t_1)$. The numbers in \eref{eq:Ms011_max_WoAcc} are obtained by fitting the numerical result from  solving the angular momentum and energy conservation equations
%%%%%=====
\begin{subequations}\label{eq:conservation}
	\begin{align}
		\M[(011)]{a}{*c} M_1^2 - a_{*0} M_0^2 & = \frac{\M[(011)]{M}{s,1}}{\omega_{011}(t_1)} - \frac{\M[(011)]{M}{s,0}}{\omega_{011}(t_0)},\\
		M_0 + \M[(011)]{M}{s,0} & = M_1 + \M[(011)]{M}{s,1}, \label{eq:conservation_M}
	\end{align}
\end{subequations}
%%%%%
where $M_1 \equiv M(t_1)$ is the BH mass at time $t_1$.

With $\M[(011)]{M}{s,1}$ obtained in Eq.~\eqref{eq:Ms011_max_WoAcc}, the $t_1$ could be estimated as
%%%%%=====
\begin{align} \label{eq:tau_SR}
	t_1 \approx\tau_\text{SR}^{(011)}\log\frac{M_\mathrm{s,1}^{(011)}}{M_\mathrm{s,0}^{(011)}},
\end{align}
%%%%%
where $\tau_\text{SR}^{(011)}=1/(2\Gamma_{011})$ is the superradiance time scale. The superradiance rate $\Gamma_{011}$ has a simple form in the $M\mu\ll 1$ limit
%%%%%=====
\begin{align}\label{eq:Gamma011}
	M\Gamma_{011} \approx \frac{1}{48}\left(a_{*}-4M\mu\right)(M\mu)^9.
\end{align}
%%%%%
If satisfied with an order of magnitude estimate, one could further take the logarithm of the mass ratio in Eq.~\eqref{eq:tau_SR} to be roughly 10, and obtain $t_1$ in the SI units
%%%%%=====
\begin{align}\label{eq:t1_no_accretion}
	t_1 \approx 5.09\,\mathrm{yr} \left(\frac{10M_\odot}{M_0}\right)^8\left(\frac{10^{-12}\,\mathrm{eV}}{\mu}\right)^9\left(\frac{1}{a_{*0}}\right).
\end{align}
%%%%%

The attractor phase of the subdominant mode starts at $t_1$ and ends at $t_2$ when the $\{0,1,1\}$ mode reaches its maximum mass. It is an attractor of the BH since its mass is almost a constant $M_1$ and its spin decreases slightly from $a_{*\text{c}}^{(111)}$ to $a_{*\text{c}}^{(011)}$. In this time range, the $\{1,1,1\}$ mode quickly shrinks, with most of its energy falling into the horizon. In contrast, the $\{0,1,1\}$ mode continues to grow, but at a much slower rate. The masses of these two modes approximately satisfy
%%%%%=====
\begin{align}
	\M[(011)]{M}{s}\Gamma_{011}+\M[(111)]{M}{s}\Gamma_{111}=0.
\end{align}
%%%%%
During this period, the superradiant flux of the $\{1,1,1\}$ mode can be estimated by
%%%%%=====
\begin{align}\label{eq:ESR_111}
	\dE[(111)]{E}{SR}=2\M[(111)]{M}{s}\Gamma_{111} \approx -\frac{80}{19683} \frac{\M[(111)]{M}{s}}{M_1}(M_1\mu)^{12},
\end{align}
%%%%%
while the GW emission flux of the $\{1,1,1\}$ mode is
%%%%%=====
\begin{align}
	\dE[(111)]{E}{GW} &\approx \frac{2 \left(484+9 \pi ^2\right)}{32805}\frac{\M[(111)]{M}{s}}{M_1}a_{*0}(M_1\mu)^{15},
\end{align}
%%%%%
which is suppressed by $(M\mu)^3$ compared to $\dot{E}_\text{SR}$. Thus one could safely ignore the GW emission and use Eq.~\eqref{eq:ESR_111} to solve Eq.~\eqref{eq:Ms_ODE}. This observation applies to all $l>1$ modes. Then the scale of the $\{1,1,1\}$ mode lifetime could be derived as
%%%%%=====
\begin{align}\label{eq:111_life}
	{\M[(111)]{\tau}{life} }&= \frac{19683}{80}M_1 (M_1\mu)^{-12}.
\end{align}
%%%%%
The $t_2$ can then be estimated with $t_1+\tau_\text{life}^{(111)}$. In this phase, the  $\{0,1,1\}$ mode mass increases gradually from $\M[(011)]{M}{s,1}$ to $M_\text{s}^{(011)}(t_2)\approx \M[(011)]{M}{s,1}+\M[(111)]{M}{s,1}$.

The attractor phase of the dominant mode starts from $t_2$. In this phase, the BH has constant mass $M_1$ and spin $a_{*\text{c}}^{(011)}$. The GW emission dominates the evolution of the $\{0,1,1\}$ mode. One could then solve Eq.~\eqref{eq:Ms_ODE} directly and obtain
%%%%%=====
\begin{align}\label{eq:Ms_GW_011_WoAcc}
	\M[(011)]{M}{s}(t) = \frac{\M[(011)]{M}{s,2}}{1+(t-t_2)/\M[(011)]{\tau}{GW}},
\end{align}
%%%%%
where the GW emission timescale is given by
%%%%%=====
\begin{align}\label{eq:tau_GW_011_WoAcc}
	{\M[(011)]{\tau}{GW} } & = 40.22 \cdot \frac{M_1^2}{\M[(011)]{M}{s,2}}(M_1\mu)^{-14}.
\end{align}
%%%%%
In the SI units, this time scale can be estimated with
%%%%%=====
\begin{align}
	\tau_\mathrm{GW}^{(011)}\approx 4.85\times 10^6 \mathrm{yr} \left(\frac{10 M_\odot}{M_0}\right)^{14} \left(\frac{10^{-12}\mathrm{eV}}{\mu}\right)^{15} \left(\frac{1}{a_{*0}}\right).
\end{align}
%%%%%

In a realistic case, the $l=m=2$ modes become important when the $\{0,1,1\}$ mode is almost depleted via GW emission. The latter evolution is very similar to the process with $l=m=1$ modes described above. The analytic estimates are still valid for the $l=m=2$ modes, with the initial BH spin and mass now replaced by $\M[(011)]{a}{*c}$ and $M_1$, respectively. More details can be found in Ref.~\cite{Guo:2022mpr}.

\section{Effect of accretion on single mode evolution}\label{sec:single}

In this section, we study the effect of accretion on the mode $\{0,1,1\}$, which has the largest superradiance rate. Other modes are ignored at the moment. The important time scales are analyzed in Sec.~\ref{sec:time_scales}. If the superradiance could happen initially without accretion, the evolution further depends on the relative size of these time scales, which are explained below in Sec.~\ref{sec:acc_ll_SR}-\ref{sec:acc_g_GW}. If the superradiance is forbidden initially, the evolution is a little different, which is studied in Sec.~\ref{sec:acc_forbiden}. Finally, we summarize this section and point out the universal behavior of the BH evolution in Sec.~\ref{sec:single-diss}.

\subsection{Time Scales}\label{sec:time_scales}

If superradiance could happen without accretion, there are two widely separated scales: the superradiance time scale $\tau_\text{SR}$ and the GW emission time scale $\tau_\text{GW}$, with the latter much larger than the former. With accretion, there is a new time scale $\tau_\text{acc}$. Another time scale can be found by requiring the critical BH spin in Eq.~\eqref{eq:a_critc} to be less than the limiting value 0.998, which gives a critical value of the BH mass
%===
\begin{align}\label{eq:M_c}
	M_\text{de}^{(nlm)} \approx \frac{0.47m}{\mu},
\end{align}
%===
where the subscript stands for ``decaying". Once the BH mass surpasses this critical value, the superradiant mode turns to a decaying mode. In this section, we consider the case with $M_0<M_\text{de}^{(011)}$. We refer to the time cost as the decaying time scale $\tau_\text{de}$. Using Eq.~\eqref{eq:BH_mass_efold}, one could estimate this decaying time scale as
%===
\begin{align}
\tau_\text{de}=  \tau_\text{acc}\log\frac{0.47m}{\alpha_0},
\end{align}
%===
where $\alpha_0=M_0\mu$. For phenomenologically important cases, the value of $\alpha_0$ is roughly $0.01$, making the logarithm of order $\mathcal{O}(1)$. Thus, $\tau_\text{de}$ is of the same order as $\tau_\text{acc}$ and is not a separate time scale. 

\begin{figure*}[tbp]
	\centering 
	\includegraphics[width=0.9\textwidth]{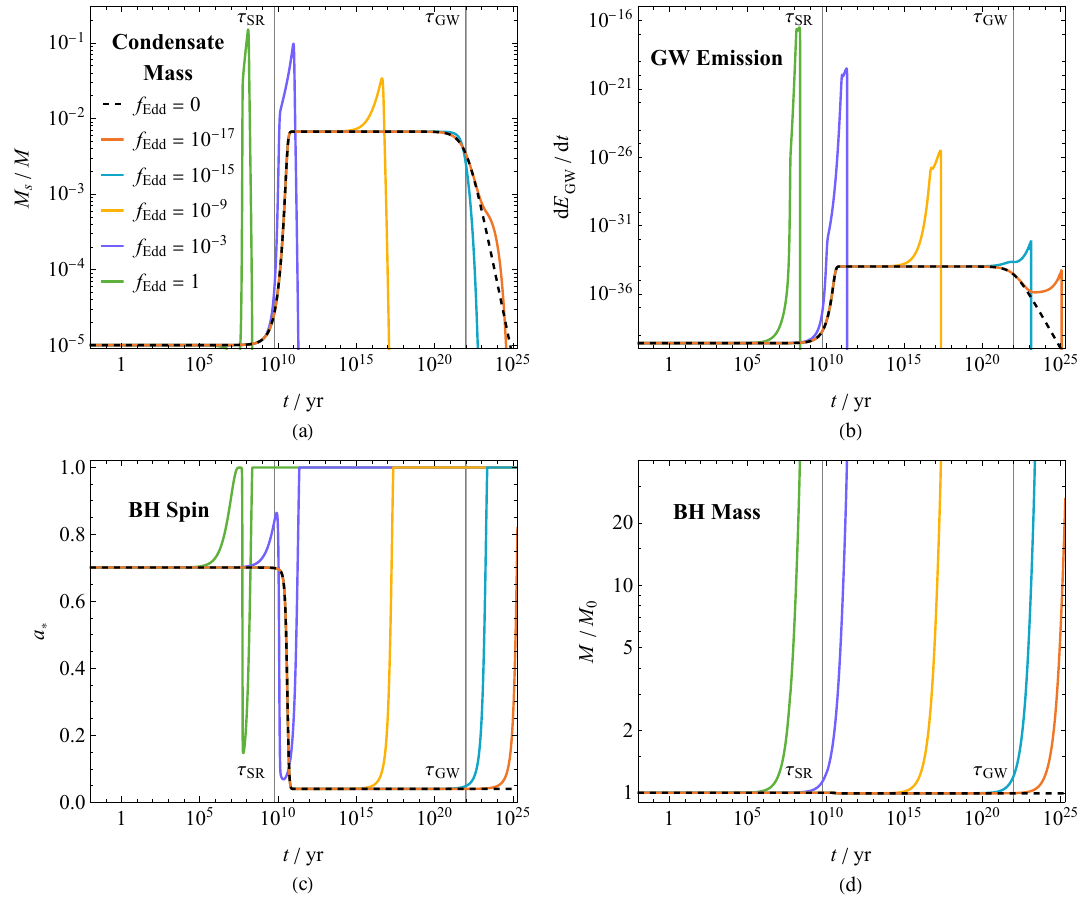} 
	\caption{The time evolution of scalar condensate masses (panel a), GW emission fluxes (panel b), BH spins (panel c), and BH mass (panel d). Initial parameters are $a_{*0}=0.7$, $M_0=10^3M_\odot$, $\M[(011)]{M}{s,0}=10^{-5}M_0$ and $M_0\mu=0.01$. Different colors represent different accretion rates indicated by the legends. The vertical lines are the superradiant timescale $\tau_\text{SR}$ and the GW emission timescale $\tau_\text{GW}$.} 
	\label{fig:vs_fEdd} 
\end{figure*}

The value of $\tau_\text{acc}$ can be adjusted by changing the accretion rate $f_\text{Edd}$ as in Eq.~\eqref{eq:tau_acc}. There are five scenarios with this new scale, i,e, $\M[]{\tau}{acc}\ll\M[(011)]{\tau}{SR}$, $\M[]{\tau}{acc}\sim\M[(011)]{\tau}{SR}$, $\M[(011)]{\tau}{SR}\ll\M[]{\tau}{acc}\ll\M[(011)]{\tau}{GW}$, $\M[]{\tau}{acc}\sim\M[(011)]{\tau}{GW}$ and $\M[]{\tau}{acc}\gg\M[(011)]{\tau}{GW}$. In Fig.~\ref{fig:vs_fEdd}, we numerically solve Eqs.~\eqref{eq:total_ODEs} for these five scenarios. The initial BH mass and spin are chosen as $M_0=10^3M_\odot$ and $a_{*0}=0.7$, respectively. The initial mass coupling is set as $M_0\mu=0.01$, which then gives the scalar mass $\mu\approx 1.34\times10^{-15}\,\mathrm{eV}$. The superradiance and the GW emission time scales without accretion are $\M[(011)]{\tau}{SR}\approx6.01\times10^{9}\,\mathrm{yr}$ and $\M[(011)]{\tau}{GW}\approx1.03\times10^{22}\,\mathrm{yr}$. Their values are plotted in Fig.~\ref{fig:vs_fEdd} as vertical lines. The case without accretion ($f_\text{Edd}=0$) is plotted in dashed curves for comparison. One can see that the BH-condensate system evolution is dramatically affected by a large accretion rate. Specifically, both the superradiance and the GW emission proceed much faster than the case with $f_\text{Edd}=0$. By observing the curves in Fig.~\ref{fig:vs_fEdd}, we further categorize the above five scenarios into three groups, which are studied in the next three subsections respectively.

On the other hand, if the superradiance could not happen initially, neither the superradiance nor the GW emission time scales exist. The BH evolution is controlled by the accretion until the extraction of the angular momentum is faster than the inwards flow due to the accretion. This case is studied in Sec.~\ref{sec:acc_forbiden}.

%%%%%FFFFF
%%%%%

\subsection{\texorpdfstring{$a_{*0}>a_{*\text{c}}^{(001)}$ and $\M[]{\tau}{acc}\ll\M[(011)]{\tau}{SR}$}{acc << SR}}
\label{sec:acc_ll_SR}

\begin{figure*}[htbp]
	\centering 
	\includegraphics[width=0.9\textwidth]{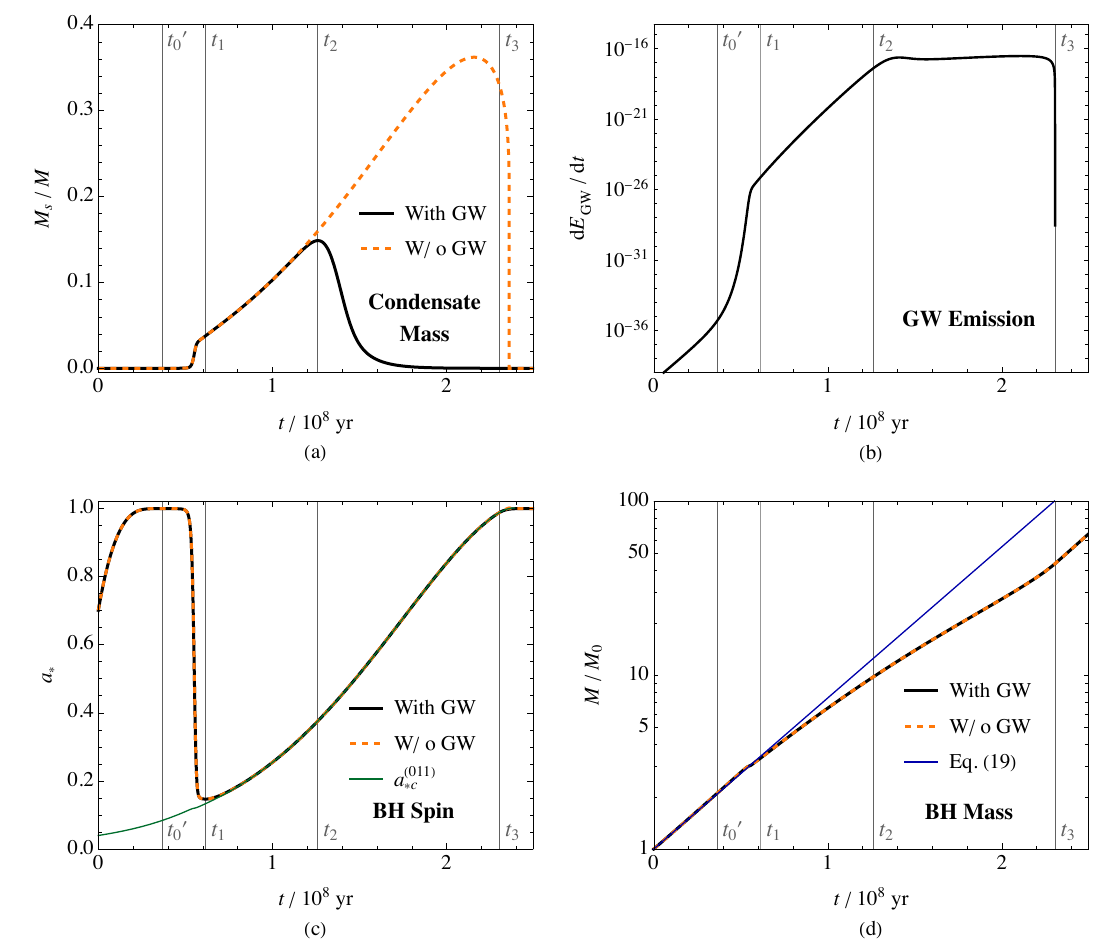}
	\caption{The time evolution of the scalar condensate mass (panel a), the GW emission (panel b), the BH spin (panel c), and the BH mass (panel d). The black and orange dotted curves are the evolution with and without the GW emission, respectively. The green curve in panel (c) is calculated with \eref{eq:a_critc}. The blue line in panel (d) is the exponential increase of the BH mass given in \eref{eq:BH_mass_efold}.} 
	\label{fig:single} 
\end{figure*}

The case of $\M[]{\tau}{acc}\ll\M[(011)]{\tau}{SR}$ corresponds to the curves with $\M[]{f}{Edd}=1$ in Fig.~\ref{fig:vs_fEdd}. They are re-plotted with more details in \fref{fig:single}. The evolution was split into four distinct phases in \rcite{Guo:2022mpr}. Below we briefly review these phases:\footnote{Here we rename the phases of those in \rcite{Guo:2022mpr}.}

%%%%%FFFFF
%%%%%

%%%
\begin{itemize}
	\item {\it Spin-up phase}. 
	This phase corresponds to the interval $0<t<t_0^\prime$, where $t_0^\prime$ represents the time at which the BH spin reaches its first local maximum. During this phase, the BH evolves following the process described in Sec.~\ref{sec:evo_wo_scalar}, while the scalar condensate is too small in mass to be important.
	\item {\it Spin-down phase}. 
	This phase spans $t_0^\prime<t<t_1$, where $t_1$ marks the time at which the BH spin reaches its minimum. As the BH mass and the condensate mass increase with time, the angular momentum extraction rate grows faster than exponentially. At some point, the extraction surpasses the accretion, resulting in a decrease of $a_{*}$ until it approaches the $\{0,1,1\}$ Regge trajectory determined by \eref{eq:a_critc}. This Regge trajectory is plotted as the green curve in Fig.~\ref{fig:single}(c).
	\item {\it Attractor phase}.
	This phase extends from $t_1$ to $t_3$, where $t_3$ denotes the moment when the BH mass reaches $M_\text{de}^{(011)}$ given in Eq.~\eqref{eq:M_c}. Throughout this phase, the BH evolves along the Regge trajectory of the $\{0,1,1\}$ mode, which acts as an attractor mathematically. Concurrently, as the condensate accumulates mass, its GW emission intensifies, and $\M[(011)]{M}{s}/M$ reaches its maximum at $t_2$. Note that it does not coincide with the peak of the GW emission flux in panel (b), since the latter depends on both the mass ratio and the mass coupling, illustrated in Eq.~\eqref{eq:dot_E_GW}.
	\item {\it Decaying phase}. 
	This phase lasts from $t_3$ until the condensate depletes. At $t_3$, the superradiance rate equals zero and the condensate becomes a decaying mode beyond $t_3$. Consequently, the scalar condensate quickly shrinks, returning its mass and angular momentum to the BH. For the single-mode example, $t_3$ is the time when the BH mass reaches its critical value $M_\text{de}^{(011)}$.
\end{itemize}
%%%

The important quantities in the numerical evolution can be estimated with simple analytic expressions at $\alpha\ll 1$ limit. To facilitate the analysis, we also solve the evolution equations without GW emission. The results are shown with orange dashed curves in \fref{fig:single}. The exponential rise of the BH mass in Eq.~\eqref{eq:BH_mass_efold} is also displayed as a thin blue line in Fig.~\ref{fig:single}~(d). For compactness, the quantities at a specific time are labeled with the corresponding subscripts, with subscript 0 reserved for the initial time, e.g. $\alpha_0\equiv M_0\mu$, $M_1\equiv M(t_1)$ and $M_{\text{s,0}'}\equiv M_\text{s}(t_0')$.

The spin-up phase spans from $t=0$ to $t_0'$, when the BH spin reaches the local maximum. In the case of $\M[]{\tau}{acc}\ll\M[(011)]{\tau}{SR}$, this phase can be further divided into two segments, separated by the time $t_0''$ when the BH mass reaches $\sqrt{6}kM_0$. This time can be readily calculated
%%%%%=====
\begin{align}
	t_0''=\tau_\mathrm{acc}\log\left(\sqrt{6}k\right).
\end{align}
%%%%%

In the range from $t=0$ to $t_0''$, the BH mass grows according to \eref{eq:BH_mass_efold} until $M = \sqrt{6} k M_0$, and the BH spin follows \eref{eq:as_ini_not_0} up to $a_* \approx a_{*\lim}$.
Ignoring GW emission and inserting \eref{eq:Gamma011} and \eref{eq:as_ini_not_0} into \eref{eq:Ms_ODE}, one could calculate the condensate mass
%%%%%=====
\begin{align}\label{eq:Ms011_x_l_sqrt6}
	M^{(011)}_\text{s} = M^{(011)}_\text{s0} \exp\left\{{\alpha_0^8\mu\tau_\text{acc}}\left[g\left(\frac{M}{M_0}\right) - g\left(1\right)\right]\right\}, 
	\text{ for }  1 \leq \frac{M}{M_0} \leq {\sqrt{6}\,k},&
\end{align}
%%%%%
with
%%%%%=====
\begin{align}
	g(x) \equiv & \ \frac{1}{36} \left[-\frac{2 \alpha _0 x^9}{3} +\frac{2}{7} \sqrt{6} k x^7 
		+\frac{\sqrt{3} k}{35} \left(216 k^4+36 k^2 x^2+5 x^4\right) (9 k^2-x^2)^{3/2}\right].
\end{align}
%%%%%
This estimate also applies to the case when $a_{*0}<a_{*\text{c}}^{(001)}$.

Subsequently, the BH spin remains at approximately $a_{*\lim}$ from $t_0''$ to $t_{0}' $. As the condensate mass increases, the BH spin evolution is increasingly affected by superradiance, which results in $\dot{a}_* = 0$ at $t_{0'}$. In this time range, one could simplify $\Gamma_{011}$ in Eq.~\eqref{eq:Gamma011} by taking the leading order (LO) of the $\alpha$ and setting $a_*=1$. Then \eref{eq:Ms_ODE} can be solved by ignoring the GW emission, arriving at
%===
\begin{align}\label{eq:Ms0p}
	M^{(011)}_{\text{s}} =
	M_{\text{s,0}''}  \exp\left[\frac{\mu\,\tau_\text{acc}}{192}\left(\alpha^8-\alpha_{0''}^8\right)\right],
\end{align}
%===
where $\alpha_{0''}=\sqrt{6}k\alpha_0$ and $M_{\text{s,0}''}$ is determined by \eref{eq:Ms011_x_l_sqrt6}.

The spin-down phase starts at $t_0'$ and ends at $t_1$, when the BH spin reaches the local minimum. In this time range, the accretion with photon capture first keeps the BH spin close to $a_{*\text{lim}}$ while both the BH mass and the scalar condensate mass increase with time. Eventually, close to $t_1$, the condensate is dense enough to quickly extract the energy and angular momentum from the BH, causing a rapid drop of the BH spin to $a_{*\text{c}}^{(011)}$. The GW is still irrelevant in this time range, as shown in Fig.~\ref{fig:single}~(a).

The evolution of the BH spin $a_*$ can be derived from Eqs.~\eqref{eq:M_BH_ODE} and \eqref{eq:J_BH_ODE}
%===
\begin{align}\label{eq:dasdt}
	\dot{a}_* = \frac{1}{\tau_\text{acc}}\left(\frac{J_\mathrm{ms}^{\dagger}}{M E_\mathrm{ms}^{\dagger}}-2a_*\right)
	-\frac{2M_\text{s}}{M}\left(\frac{1}{M\omega_{011}}-2a_*\right)\Gamma_{011},
\end{align}
%===
where the photon capture is ignored. From $t_0''$ to $t_0'$, both terms on the right side are small, resulting in an almost constant $a_*$. From $t_0'$ to $t_1$, the accretion term is still small, while the superradiance term grows quickly, leading to the fast drop of the BH spin. Thus one could ignore the accretion term in the time range between $t_0''$ and $t_1$. The BH mass $M$ can be estimated with Eq.~\eqref{eq:BH_mass_efold} and the condensate mass continues to evolve according to the form given by \eref{eq:Ms0p}. Then, using $\Gamma_{011}$ at the LO of the $\alpha$ and setting $a_*=1$, one rewrites Eq.~\eqref{eq:dasdt} as
%===
\begin{align}\label{eq:dasdt_term2}
	\dot{a}_* = - \frac{M_\mathrm{s,0''}^{(011)}}{24} M^6 \mu^8 \exp\left[\frac{\tau_\text{acc}}{192}(M^8-M^8_{0''})\mu^9\right].
\end{align}
%===
Integrating both sides of \eref{eq:dasdt_term2} from $t_{0''}$, one obtains the following solution
%%%%%=====
\begin{align}
	\begin{split}
		a_*(\alpha) = \, & a_{*0''} -  \frac{1}{192} M^{(011)}_\mathrm{s0''} \tau_\mathrm{acc} \mu^2  e^{-\frac{\alpha_{0''}^8 \tau_\mathrm{acc} \mu }{192}} 
		\\
		& \times \left[\alpha_{0''}^6 E_{\frac{1}{4}}\left(\frac{-\alpha_{0''}^8 \tau_\mathrm{acc} \mu }{192}\right) - \alpha^6 E_{\frac{1}{4}}\left(\frac{-\alpha^8 \tau_\mathrm{acc} \mu}{192}\right)\right],
	\end{split}
\end{align}
%%%%%
where $E_{\frac{1}{4}}(x)$ denotes the exponential integral function. At $t_1$, the BH spin decreases to a value close to $a_\mathrm{*c}^{(011)} \sim 4\alpha_1 \ll a_{*\lim}$.  Thus, one can approximate $a_{*1} - a_{*0''}$ by $-a_{*\lim}$, and perform the following expansion
%===
\begin{align}
	E_{\frac{1}{4}}\left(\frac{-\alpha_1^8 \tau_\mathrm{acc} \mu}{192}\right) = 
	e^{\frac{\alpha_1^8 \tau_\mathrm{acc} \mu }{192}} & \left\{-\frac{192}{\alpha_1^8 \tau_\mathrm{acc} \mu }+\mathcal{O}\left[\left(\frac{1}{\alpha_1^{8}\tau_\mathrm{acc}\mu}\right)^{2}\right]\right\}.
\end{align}
%===
Finally, we arrive at an estimate for $\alpha_1$
%%%%%=====
\begin{align}\label{eq:alpha_1}
	\alpha_1 = \left\{ \frac{-48}{\tau_\text{acc}\mu}W_{-1}\left[-\left(\frac{32\cdot3^{3/4}\mu M_{\mathrm{s,0''}}^{(011)}\left(\tau_{\mathrm{acc}}\mu\right)^{1/4}}{\alpha_{0''}^{6}M_{\mathrm{s,0''}}^{(011)}\tau_{\mathrm{acc}}\mu^{2}E_{\frac{1}{4}}\left(\frac{-\alpha_{0''}^{8}\tau_{\mathrm{acc}}\mu}{192}\right)-192a_{*0''}e^{\frac{\alpha_{0''}^{8}\tau_{\mathrm{acc}}\mu}{192}}}\right)^{4}\right]\right\} ^{1/8},
\end{align}
%%%%%
where $W_{-1}(x)$ is the Lambert $W$ function. If $M_{\text{s},0''}^{(011)}\mu\ll \alpha_{0''}^2$, the term with the Exponential integral function is a small contribution and can be dropped. With the value of $\alpha_1$, the time $t_1$ can be obtained using \eref{eq:BH_mass_efold} and the condensate mass is estimated with Eq.~\eqref{eq:Ms0p}.

The attractor phase spans from $t_1$ to $t_3$, when the BH mass reaches the critical value $M_\text{de}^{(011)}$ defined in Eq.~\eqref{eq:M_c}. In this time range, the BH follows the Regge trajectory closely. We further separate this phase into two ranges with $t_2$. The condensate mass to BH mass ratio first increases with time, reaches the maximum at $t_2$, and then drops gradually due to the GW emission.

From $t_1$ to $t_2$, the photon absorption from the accretion disk is unimportant since the BH spin is smaller than $0.97$. The GW radiation has a small effect on $M_s/M$ at $t_2$, which can be seen from the difference of the black and red dashed curves at $t_2$ in Fig.~\ref{fig:single}. Below we treat this effect as perturbation. At the zeroth order, the GW radiation is ignored and the  BH spin evolves along the Regge trajectory given by Eq.~\eqref{eq:a_critc}.  Inserting Eqs.~\eqref{eq:M_BH_ODE} and \eqref{eq:J_BH_ODE} into the derivative of $a_{*\text{c}}^{(011)}$ with respect to $t$, one could then solve the superradiance flux order by order in $\alpha=M\mu$
%%%%%=====
\begin{align}\label{eq:Edot_SR}
	\dE[(011)]{E}{SR} = \frac{\alpha  M}{\M[]{\tau}{acc}} \left( 3\sqrt{\frac32}-\frac{31\alpha}2+\frac{787\alpha^2}{8\sqrt{6}}-\frac{20305\alpha^3}{216} +\frac{1564549\alpha^4}{3456\sqrt{6}}+\mathcal{O}(\alpha^5)  \right).
\end{align}
%%%%%
Ignoring GW emission and using \eref{eq:M_BH_ODE}, one could integrate this equation directly
%%%%%=====
\begin{align}\label{eq:Ms011_Regge}
	{\M[(011)]{M}{s,Regge}(\alpha)} \equiv \frac{\alpha^2}{\mu}\left( \frac32\sqrt{\frac32}-\frac{2\alpha}3-\frac{473\alpha^2}{32\sqrt{6}}-\frac{223\alpha^3}{270} +\frac{999421\alpha^4}{20736\sqrt{6}}+\mathcal{O}(\alpha^5) \right).
\end{align}
%%%%%
With this expression, the LO condensate mass at time $t_2$ is
%%%%%=====
\begin{align}\label{eq:bar_Ms011_t2}
	{\M[(011),\text{LO}]{M}{s,2}} = \M[(011)]{M}{s,1}+ \M[(011)]{M}{s,Regge}(\alpha_2) - \M[(011)]{M}{s,Regge}(\alpha_1).
\end{align}
%%%%%
To obtain a more precise value of $\M[(011)]{M}{s,2}$, we next add the GW emission effect perturbatively. Assuming that the BH mass increases exponentially, the mass correction can be estimated by
%%%%%=====
\begin{align}
	\delta {\M[(011)]{ M}{s,2}} &=  -\int^{t_2}_{t_1}\dE[(011)]{E}{GW} (t)dt\approx - 5.24\times 10^{-3} {\M[]{\tau}{acc}} \alpha_2^{16},
\end{align}
%%%%%
where Eqs.~\eqref{eq:BH_mass_efold} and \eqref{eq:Ms011_Regge} have been used. Then the condensate mass at $t_2$ is
%%%%%=====
\begin{align}\label{eq:Ms011_2}
	{\M[(011)]{ M}{s,2}} &= {\M[(011),\text{LO}]{M}{s,2}} + \delta {\M[(011)]{ M}{s,2}}. 
\end{align}
%%%%%

To determine the value of $\alpha_2$, we make use of $d\left(\M[(011)]{M}{s}/M\right)/dt=0$ at $t_2$, which gives
%%%%%=====
\begin{align}\label{eq:for_alpha_2}
	\left(1+\frac{\M[(011)]{M}{s,Regge}(\alpha_2)}{M_2}\right)\dE[(011)]{E}{SR}-\dE[(011)]{E}{GW} = \frac{\M[(011)]{M}{s,Regge}(\alpha_2)}{\M[]{\tau}{acc}}.
\end{align}
%%%%%
Here we have assumed the condensate obtains most of its mass in the attractor phase, i.e., $M_\text{s}^{(011)}(\alpha_2)\approx M_\text{s,Regge}^{(011)}(\alpha_2)$. Inserting the expressions in Eqs.~\eqref{eq:Edot_SR}, \eqref{eq:Ms011_Regge} and \eqref{eq:dot_E_GW} and assuming $\alpha_2\gg \alpha_1$, one finds the perturbation is invalid unless $\mu\tau_\text{acc}\sim \alpha_2^{14}$. Then the LO value of $\alpha_2$ is
%%%%%=====
\begin{align} \label{eq:alpha_2}
	\alpha_2 \approx 1.25 \times \sqrt[14]{\frac{1}{\mu \M[]{\tau}{acc} }}.
\end{align}
%%%%%

The value of $t_2$ can be estimated with \eref{eq:M_BH_ODE}. Ignoring the photon-capturing and using the expression in Eq.~\eqref{eq:Edot_SR}, one arrives at
%%%%%=====
\begin{align}\label{eq:dot_M_corr}
	\dot{M}  \approx \frac{M}{\M[]{\tau}{acc}}  - \dE[(011)]{E}{SR} \approx \frac{M}{\M[]{\tau}{acc}}  - 3\sqrt{\frac32} \frac{M^2\mu}{ \M[]{\tau}{acc}},
\end{align}
%%%%%
at the LO. This differential equation can be integrated directly, which gives
%%%%%=====
\begin{align}\label{eq:M_g_t_1}
	M  \approx \frac{2 M_1 e^{(t-t_1)/\tau }}{2 + 3 \sqrt{6} M_1 \mu \left[e^{(t-t_1)/\tau } - 1\right]}.
\end{align}
%%%%%
Then $t_2$ could be obtained as
%%%%%=====
\begin{align}\label{eq:t_2}
	t_2  \approx t_1 + \M[]{\tau}{acc}\log\left[\frac{\alpha_2(2-3\sqrt{6}\,\alpha_1)}{\alpha_1(2-3\sqrt{6}\,\alpha_2)}\right] .
\end{align}
%%%%%
With the initial parameters used in Fig.~\ref{fig:single}, the estimates of $\M[(011)]{M}{s,2}/M_2$, $\alpha_2$ and $t_2$ are compared to the numerical results in Tab.~\ref{tab:single}.

The second part of the attractor phase starts at $t_2$ and ends at $t_3$. In this time range, GW emission begins to dominate the evolution of the condensate. According to \eref{eq:dot_E_GW}, the GW emission flux is determined by both the condensate mass $\M[(011)]{M}{s}$ and the BH mass $M$.  From $t_2$ to $t_3$, the former decreases over time while the latter grows almost exponentially. The resulting GW flux could increase, decrease or even be a constant, depending on the initial parameters. For the parameters used in \fref{fig:single}, the GW emission flux is slightly enhanced in this time range. By definition, the corresponding BH mass $M_3\approx M_\text{de}^{(011)}$. The LO expression of $M_\text{de}^{(011)}$ is given in Eq.~\eqref{eq:M_c}. Higher-order contributions of $M_\text{de}^{(011)}$ could be solved from
%%%%%=====
\begin{align}\label{eq:for_alpha_3}
	\M[]{a}{*lim} = \frac{4 M_3 \omega_{011}}{1 + 4 M_3^2 \omega_{011}^2}, 
\end{align}
%%%%%
where $\omega_{011}$ can be found in Eq.~\eqref{eq:omega} with $\alpha$ replaced by $\alpha_3$. The obtained mass coupling is $\alpha_3=M_3\mu\approx 0.469$. At $t_3$, Eq.~\eqref{eq:dot_M_corr} is not a good estimate due to the ignorance of the higher order terms. According to Eqs.~\eqref{eq:M_BH_ODE} and \eqref{eq:Ms_ODE}, if GW emission and photon capturing are ignored, the sum of the BH mass and the condensate mass increases almost exponentially. Then $t_3$ can be estimated by
%%%%%=====
\begin{align}\label{eq:t_3}
	t_3 &= t_2 + \M[]{\tau}{acc}\log\left[\frac{M_3+\M[(011)]{M}{s,Regge}(\alpha_3)}{M_2+\M[(011)]{M}{s,Regge}(\alpha_2)}\right].
\end{align}
%%%%%

Finally, in the decaying phase beyond $t_3$, the condensate is quickly absorbed into the BH.

%%%%%TTTTT
\begin{table*}[t]\footnotesize
	\centering
	\renewcommand\arraystretch{1.2}
	\begin{tabular}{p{0.4cm}<{\centering}p{0.5cm}<{\centering}p{1.52cm}<{\centering}p{1.52cm}<{\centering}p{1.45cm}<{\centering}p{1.52cm}<{\centering}p{1.52cm}<{\centering}p{1.45cm}<{\centering}p{1.45cm}<{\centering}}
	\hline
	\hline
	\multicolumn{2}{c}{$\M[]{f}{Edd}$} & $\M[(011)]{M}{s,1}/M_1$ & $\alpha_1$ & $t_1$ / yr & $\M[(011)]{M}{s,2}/M_2$ & $\alpha_2$ & $t_2$ / yr & $t_3$ / yr \\
	\hline
	\multirow{2}{*}{1}  & Est. & $2.93 \times 10^{-2}$ & $2.93 \times 10^{-2}$ & $5.38 \times 10^{7}$ & $1.49 \times 10^{-1}$ & $9.73 \times 10^{-2}$ & $1.30 \times 10^{8}$ & $2.21 \times 10^{8}$\\
	 & Num. & $3.78 \times 10^{-2}$ & $3.35 \times 10^{-2}$ & $6.16 \times 10^{7}$ & $1.49 \times 10^{-1}$ & $9.75 \times 10^{-2}$  & $1.26 \times 10^{8}$ & $2.41 \times 10^{8}$\\
	\hline
	\multirow{2}{*}{$10^{-3}$}  & Est. & $7.43 \times 10^{-3}$ & $1.25 \times 10^{-2}$ & $1.12 \times 10^{10}$ & $9.55 \times 10^{-2}$ & $5.94 \times 10^{-2}$ & $9.91 \times 10^{10}$ & $2.18 \times 10^{11}$\\
	 & Num. & $1.87 \times 10^{-2}$ & $1.54 \times 10^{-2}$ & $2.28 \times 10^{10}$ & $9.63 \times 10^{-2}$  & $5.97 \times 10^{-2}$  & $9.83 \times 10^{10}$ & $2.33 \times 10^{11}$\\
	\hline
	\multirow{2}{*}{$10^{-9}$}  & Est. & $6.66 \times 10^{-3}$ & $9.93 \times 10^{-2}$ & $3.90 \times 10^{10}$ & $3.26 \times 10^{-2}$ & $2.21 \times 10^{-2}$ & $4.25 \times 10^{16}$ & $2.13 \times 10^{17}$\\
	 & Num. & $6.66 \times 10^{-3}$ & $9.93 \times 10^{-3}$ & $2.23 \times 10^{11}$ & $3.36 \times 10^{-2}$ & $2.30 \times 10^{-2}$  & $4.43 \times 10^{16}$ & $2.39 \times 10^{17}$\\
	\hline
	\multirow{2}{*}{$10^{-15}$}  & Est. & $6.66 \times 10^{-3}$ & $9.93 \times 10^{-2}$ & $3.90 \times 10^{10}$ & N/A & N/A & N/A & $2.12 \times 10^{23}$\\
	& Num. & $6.66 \times 10^{-3}$ & $9.93 \times 10^{-3}$ & $2.23 \times 10^{11}$ & N/A & N/A & N/A & $2.45 \times 10^{23}$\\
   	\hline
	\multirow{2}{*}{$10^{-17}$}  & Est. & $6.66 \times 10^{-3}$ & $9.93 \times 10^{-2}$ & $3.90 \times 10^{10}$ & N/A & N/A & N/A & $2.12 \times 10^{25}$\\
	 & Num. & $6.66 \times 10^{-3}$ & $9.93 \times 10^{-3}$ & $2.23 \times 10^{11}$ & N/A & N/A & N/A & $2.49 \times 10^{25}$\\
	\hline
	\multirow{2}{*}{0}  & Est. & $6.66 \times 10^{-3}$ & $9.93 \times 10^{-2}$ & $3.90 \times 10^{10}$ & N/A & N/A & N/A & N/A \\
	 & Num. & $6.66 \times 10^{-3}$ & $9.93 \times 10^{-3}$ & $2.23 \times 10^{11}$ & N/A & N/A & N/A & N/A\\
	\hline
	\end{tabular}
	\caption{Comparison of the analytical estimates to the numerical results obtained from solving Eqs.~\eqref{eq:total_ODEs} for essential quantities. Initial parameters are identical to those used in Fig.~\ref{fig:vs_fEdd}. Since the condensate mass does not present a maximum in the attractor phase for $\M[]{f}{Edd}=0$, $10^{-17}$ and $10^{-15}$, the corresponding quantities at $t_2$ are labeled with ``N/A" in the table. Moreover, $t_3$ of $\M[]{f}{Edd}=0$ case is not applicable, which is also labeled with ``N/A".}
	\label{tab:single}
\end{table*}
%%%%%%%%%%

\subsection{\texorpdfstring{$a_{*0}>a_{*\text{c}}^{(001)}$ and $\M[]{\tau}{acc}\sim\M[(011)]{\tau}{SR}$}{SR sim acc}}
\label{sec:SR_l_acc_ll_GW}

The case of $\M[]{\tau}{acc} \sim \M[(011)]{\tau}{SR}$ is represented by the curves with $\M[]{f}{Edd} = 10^{-3}$ in Fig.~\ref{fig:vs_fEdd}. This case behaves similarly to the case of $\M[]{f}{Edd} = 1$, except that the spin does not reach $a_{*\lim}$ in the spin-up phase, as shown in Fig.~\ref{fig:vs_fEdd}(c). Since the spin-down phase, attractor phase and decaying phase are the same as in the previous case, below we focus on the spin-up phase.

The spin-up phase ends at $t_0'$, when the BH spin reaches the local maximum for the first time. Expanding the first term on the right side of Eq.~\eqref{eq:dasdt} at $a_*=1$ and combining it with \eref{eq:dasdt_term2}, Eq.~\eqref{eq:dasdt} reduces to
%===
\begin{align}\label{eq:asdot_judge}
	\dot{a}_* = &\frac{3\cdot 2^{1/3}}{\tau_\text{acc}}\left(1-a_*\right)^{2/3}
	-\frac{M_\text{s,0}}{24}M^6 \mu^8 \exp\left[\frac{\tau_\text{acc}}{192}(M^8-M^8_0)\mu^9\right].
\end{align}
%===
At $t_0'$, the two terms on the right side cancel. Making approximation $1-a_* \sim 1$, one could solve
%===
\begin{align}\label{eq:alpha_0p}
	\alpha_{0'} = &\left(\frac{144}{\tau_{\mathrm{acc}}\mu}\right)^{{1}/{8}}
	\left[
	W_0\left(2^{{4}/{9}}3^{{2}/{3}} M_\text{s,0}^{-{4}/{3}} \tau_\text{acc}^{-{1}/{3}}\mu^{-{5}/{3}}e^{\frac{\alpha_{0}^8\tau_{\mathrm{acc}}\mu}{144}}
	\right)
	\right]^{{1}/{8}},
\end{align}
%===
The time $t_0'$ can then be calculated with Eq.~\eqref{eq:BH_mass_efold}, and the BH spin and condensate mass at $t_0'$ can be derived by Eqs.~\eqref{eq:as_ini_not_0} and \eqref{eq:Ms011_x_l_sqrt6}, respectively. The subsequent evolution is the same as the scenario described in Sec.~\ref{sec:acc_ll_SR}, with the subscript $0''$ in Eq.~\eqref{eq:alpha_1} replaced by $0'$.

Actually, Eq.~\eqref{eq:alpha_0p} could be used to judge which one of the three groups the evolution should be attributed to. If the obtained $\alpha_{0'}$ is larger than $\sqrt{6} k \alpha_0$, one should instead use the analysis in Sec.~\ref{sec:acc_ll_SR}. On the other hand,  if $\alpha_{0'}$ is smaller than $\alpha_0$, the evolution does not follow the accretion trajectory at all, which case is discussed in Sec.~\ref{sec:acc_g_GW} below.

\subsection{\texorpdfstring{$a_{*0}>a_{*\text{c}}^{(001)}$ and $\M[]{\tau}{acc}\gg\M[(011)]{\tau}{SR}$}{acc >> SR}}
\label{sec:acc_g_GW}

This group includes three different scenarios $\M[(011)]{\tau}{SR} \ll \M[]{\tau}{acc} \ll \M[(011)]{\tau}{GW}$, $\M[]{\tau}{acc} \sim \M[(011)]{\tau}{GW}$ and  $\M[]{\tau}{acc} \gg \M[(011)]{\tau}{GW}$, corresponding to  $\M[]{f}{Edd} = 10^{-9}$, $10^{-15}$ and $10^{-17}$ in Fig.~\ref{fig:vs_fEdd}, respectively. The curves with $\tau_\text{acc}\gg \tau_\text{SR}^{(011)}$ deviate from the $f_\text{Edd}=0$ case only from $t\sim \tau_\text{acc}$. The whole evolution can also be split into different phases. Specially, compared to the two cases analysed in Sections~\ref{sec:acc_ll_SR} and \ref{sec:SR_l_acc_ll_GW}, the spin-up phase vanishes when $\tau_\text{acc}\gg \tau_\text{SR}^{(011)}$ and $a_{*0}>a_{*\text{c}}^{(011)}$.

The BH spin drops from $t=0$ and reaches the minimum at time $t_1$. The BH mass $M_1$ and the condensate mass $M_\text{s,1}$ can be calculated in the same way as in Sec.~\ref{sec:evo_wo_acc}. After $t_1$, the BH follows the Regge trajectory. The condensate mass may either increase or decrease, depending on the relative size of the superradiance rate and the GW emission rate at $t_1$. No matter which case, Eq.~\eqref{eq:alpha_2} can be used to estimate $\alpha_2$. If $\alpha_2>\alpha_1$, the condensate continues to grow after $t_1$ and $M_\mathrm{s}^{(011)}/M$ reaches its maximum value at $t_2$. Otherwise, the condensate shrinks since $t_1$. In the case of $\alpha_2>\alpha_1$, the time $t_2$ can be further calculated using \eref{eq:t_2}, and the condensate mass at $t_2$ can be estimated with \eref{eq:Ms011_2}.

\subsection{\texorpdfstring{$a_{*0}<a_{*\text{c}}^{(011)}$}{a0<ac}}\label{sec:acc_forbiden}

\begin{figure*}[tbp]
	\centering 
	\includegraphics[width=0.9\textwidth]{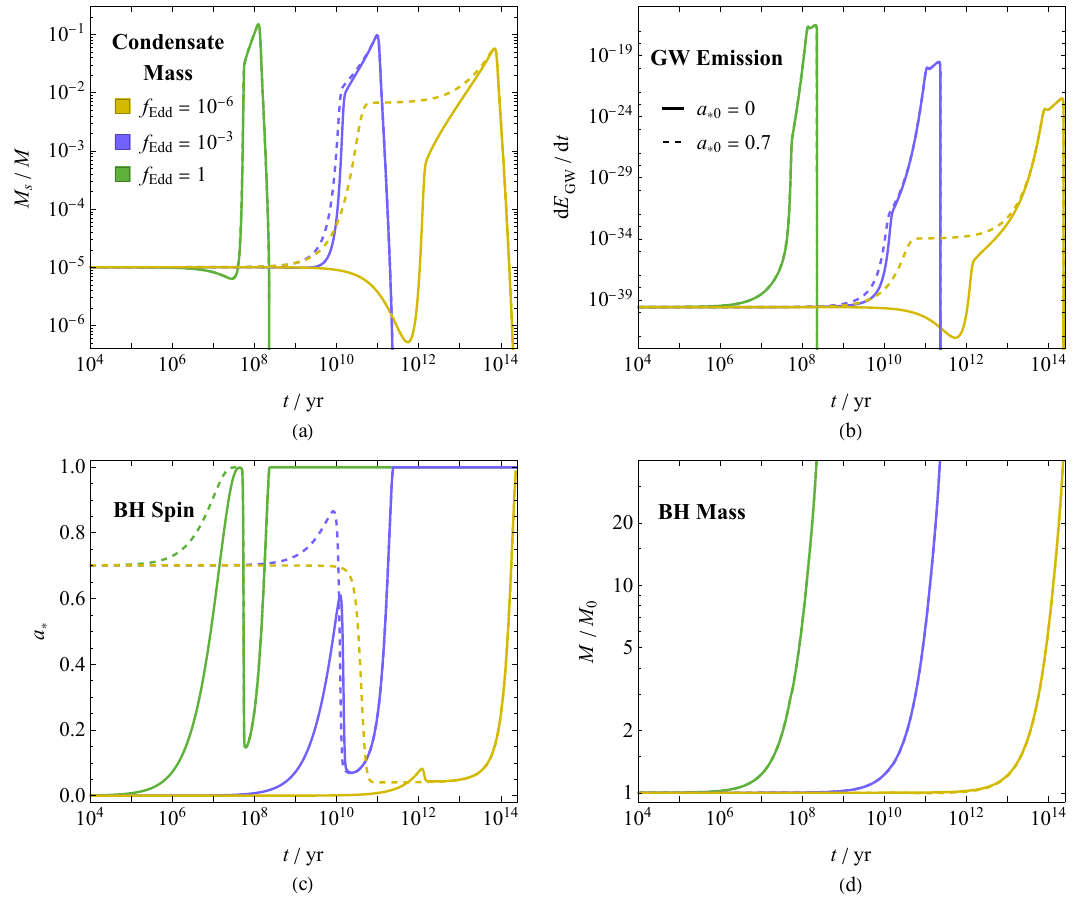} 
	\caption{The time evolution of scalar condensate masses (panel a), GW emission fluxes (panel b), BH spins (panel c), and BH masses (panel d). Initial parameters are $M_0=10^3M_\odot$, $\M[(011)]{M}{s,0}=10^{-5}M_0$, and $M_0\mu=0.01$. The solid and dashed curves are for initial BH spins $a_{*0}=0.7$ and $a_{*0}=0$, respectively. Different colors represent different accretion rates, indicated by the legends.} 
	\label{fig:vs_fEdd_2} 
\end{figure*}

When $a_{*0}$ is smaller than the critical value $a_\text{c}^{(011)}$, the superradiance condition is not satisfied initially. Nonetheless, the evolution of the BH-condensate system can still be split into the four phases explained above, as shown in Fig.~\ref{fig:vs_fEdd_2}. The angular momentum and energy flow from the condensate to the BH at first. Since the initial condensate mass is much smaller than the BH mass, this contribution is unimportant and the BH mass evolution is dominated by accretion. As a result, the growth of the BH mass can be estimated with Eq.~\eqref{eq:BH_mass_efold}. The BH spin also increases, following the relation in Eq.~\eqref{eq:as_ini_not_0}. Eventually, the BH spin is above the critical value and the superradiance is turned on. When superradiance balances accretion, the BH spin reaches its local maximum, which is the end of the spin-up phase. After the local maximum of the BH spin, the BH spin drops quickly to the Regge trajectory and then evolves along it afterward. The condensate mass first drops before the BH spin grows to its critical value $a_{*\text{c}}^{(011)}$. It can be estimated using \eref{eq:Ms011_x_l_sqrt6} during this time interval.
Since then, the system evolves as in the $a_{*0}>a_\mathrm{*c}^{(011)}$ cases, which are explained in previous subsections.

In Fig.~\ref{fig:vs_fEdd_2}, we also compared the curves with $a_{*0} =0$ to those with $a_{*0} =0.7$. Because the condensate initial mass is small, the evolutions of BH mass are indistinguishable between these two cases, confirming our argument that the BH mass growth is dominated by accretion. In contrast, the BH spin and condensate mass evolution have different behavior with different values of $a_{*0}$. Especially, the spin-up and spin-down phases are largely affected by the initial BH spin. The corresponding estimates are similar to those in Sec.~\ref{sec:SR_l_acc_ll_GW}, with a revision to the calculation of $\alpha_{0'}^{(011)}$. When $a_{*0}<a_\mathrm{*c}^{(011)}$, the spin-up phase can be divided into two segments, separated by the time $t_\mathrm{int}$ when the BH spin reaches $a_{*\text{c}}^{(011)}$. The BH mass at this time can be obtained by solving Eq.~\eqref{eq:a_critc} together with Eq.~\eqref{eq:as_ini_not_0}. The mass coupling at this time, defined as $\alpha_\text{int}$, can be accomplished either numerically or analytically in powers of $\alpha_0$
%===
\begin{align}\label{eq:intersection}
\frac{\alpha_\text{int}}{\alpha_0}=k+k\sum_{n=1}^{+\infty} c_n \left(\frac{k\alpha_0}{m}\right)^n,
\end{align}
%===
where the coefficients $c_n$ can be obtained by solving \eref{eq:as_ini_not_0} with \eref{eq:a_critc} and \eref{eq:intersection} in orders of $\alpha_0$,
%===
\begin{subequations}
\begin{align}
c_1&=2^{5/2}3^{-3/2},\\
c_2&=94/27,\\
c_3&=\frac{2^{3/2}}{3^{7/2}}\left(172-9\frac{m^2}{\bar{n}^2}\right),\\
c_4&=\frac{2}{729}\left(12451-1701\frac{m^2}{\bar{n}^2}\right).
\end{align}
\end{subequations}
%===
With the obtained $\alpha_\text{int}$, the condensate mass at $t_\text{int}$ can be estimated with \eref{eq:Ms011_x_l_sqrt6}. 
Beyond $t_\text{int}$, the formulas obtained above for the $a_{*0} > a_\text{*c}^{(011)}$ scenario can be applied. One could simply replace the $\alpha_0$, $a_{*0}$ and $M_\text{s,0}$ by $\alpha_\text{int}$, $a_{*\text{c}}^{(011)}$ and $M_\text{s}(t_\text{int})$, respectively.

It is noteworthy that if the initial condensate mass is sufficiently large, the BH spin reaches $\M[(011)]{a}{*c}$ at a faster rate with the additional contribution from the scalar field. Although this scenario can be easily adapted from Sec.~\ref{sec:acc_g_GW}, it is beyond the scope of this work. If modes with higher $l$ are considered, the BH spin may not reach $a_{*\text{c}}^{(011)}$. This scenario will be discussed in the next section.

%%%%%FFFFF
%%%%%

\subsection{Evolution on the Regge plot}\label{sec:single-diss}

%%%%%FFFFF
\begin{figure}[htbp]
	\centering 
	\includegraphics[width=0.9\textwidth]{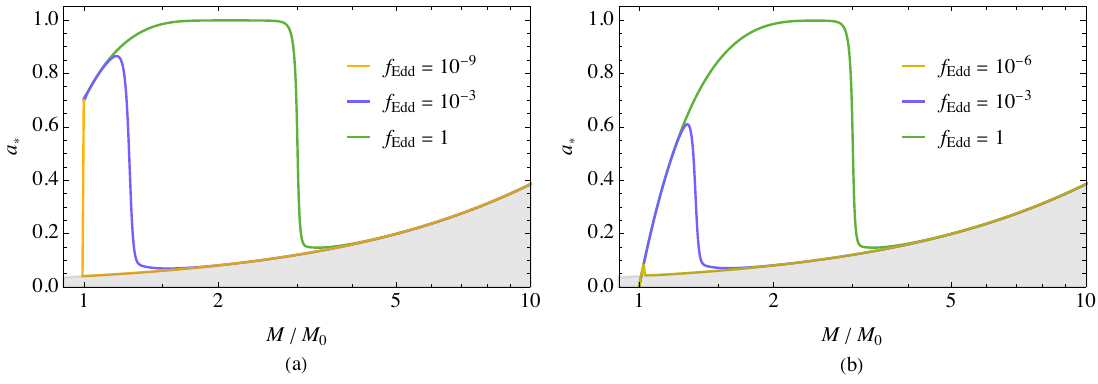}
	\caption{The evolution of the BH spin as a function of the BH mass. The initial parameters for panels (a) and (b) are identical to those used in Fig.~\ref{fig:vs_fEdd} and Fig.~\ref{fig:vs_fEdd_2}, respectively. The $\{0,1,1\}$ Regge trajectory defined by \eref{eq:a_critc} is represented as the upper boundary of the shaded region.} 
	\label{fig:Regge} 
\end{figure}
%%%%%

A classical BH free of charge has only two parameters: mass $M$ and spin $a_*$. The plot with $M$ as the horizontal axis and $a_*$ as the vertical axis is usually called {\it Regge plot} in the literature.
Despite the complications in the evolution of the BH-condensate systems with different accretion rates, the evolution presents a rather simple and universal behavior on the Regge plot. In Fig.~\ref{fig:Regge}(a), we show the evolutions of the system with three typical values of $f_\text{Edd}$, corresponding to the three scenarios detailed in Sec.~\ref{sec:acc_ll_SR}, Sec.~\ref{sec:SR_l_acc_ll_GW} and Sec.~\ref{sec:acc_g_GW}, respectively. The initial BH spin and mass are $a_{*0}=0.7$ and $M_0=10^3M_\odot$, respectively. The initial condensate mass is $M_{\text{s},0}=10^{-5}M_0$ and the initial mass coupling is $\alpha_0=0.01$. In the evolution, only the $\{0,1,1\}$ mode is considered. The GW emission of the spinning condensate is also included. In Fig.~\ref{fig:Regge}(b), the initial BH spin is $a_{*0}=0$, corresponding to the scenario detailed in Sec.~\ref{sec:acc_forbiden}. In literature, the curves defined by \eref{eq:a_critc} are commonly referred to as \textit{Regge trajectories}, which correspond to the upper boundary of the shaded region in Fig.~\ref{fig:Regge}.

We first examine the case of very slow accretion with $a_{*0} > a_{*\mathrm{c}}^{(011)}$, corresponding to $f_\text{Edd}=10^{-9}$ in Fig.~\ref{fig:Regge}(a). The superradiance dominates the evolution, leading to a fast drop of the BH spin to the $\{0,1,1\}$ Regge trajectory. The BH mass decreases slightly, with the discrepancy transferred to the condensate. Then the system evolves along the Regge trajectory.

We then turn to the case with the fastest accretion, regardless of whether $a_{*0} > a_{*\mathrm{c}}^{(011)}$, corresponding to $f_\text{Edd}=1$ in Fig.~\ref{fig:Regge}. The accretion dominates the evolution of the BH and the superradiance can be safely ignored initially. The BH first follows the trajectory determined by Eq.~\eqref{eq:das_dlnM_with_photon}, referred to as {\it accretion trajectory} below. The BH spin approaches $a_{*\text{lim}}$ and the BH mass grows almost exponentially. During this time, the condensate grows in mass and angular momentum. At some time, the condensate is dense enough such that the superradiance surpasses the accretion, leading to the fast drop to the Regge trajectory. The evolution follows the Regge trajectory afterward. 

When the accretion rate is neither large nor small, the BH follows the accretion trajectory for a while and quickly drops to the Regge trajectory. 

From the Regge plot, it is clear that the entire evolution has a universal behavior. It consists of the parts following the two trajectories and the transition between them. They correspond to the spin-up, spin-down, and attractor phases in Sec.~\ref{sec:acc_ll_SR}. The decaying phase could not be observed in the BH Regge plot. It is clear by looking at the evolution of the condensate mass. 

Due to this universal behavior, one only needs to identify the BH masse $M$ or spin $a_*$ at these critical points at which the system leaves the accretion trajectory and merges to the Regge trajectory. To determine these key points and the evolution between them, the following procedure can be applied. If $M_0\omega_{011} \geq 1 /2$, superradiance does not occur and the system follows the accretion trajectory. If $M_0\omega_{011} < 1 /2$, $\alpha_{0'}$ can be calculated using \eref{eq:alpha_0p}, as in Sec.~\ref{sec:single} and Sec.~\ref{sec:acc_forbiden}. Based on the obtained $\alpha_{0'}$, there are three possible cases:
%%%%%%
\begin{itemize}
	\item $\alpha_{0'} > \sqrt{6}k\alpha_0$: 
	This case corresponds to the scenario in Sec.~\ref{sec:acc_ll_SR} and the curves with $f_\text{Edd}=1$ in Fig.~\ref{fig:Regge}. The system initially follows the accretion trajectory approximately described by Eq.~\eqref{eq:as_ini_not_0}. Then it rapidly transits to the Regge trajectory in Eq.~\eqref{eq:a_critc} at $\alpha_{1}$ given in Eq.~\eqref{eq:alpha_1}. After that, the system follows the $\{0,1,1\}$ Regge trajectory until $M=M_\mathrm{de}^{(011)}$. 
	%%%
	\item $\alpha_{0}< \alpha_{0'} < \sqrt{6}k\alpha_0$: 
	This case corresponds to the scenario in Sec.~\ref{sec:SR_l_acc_ll_GW} and the curves with $f_\text{Edd}=10^{-3}$ in Fig.~\ref{fig:Regge}.	The system first follows the accretion trajectory and leaves this trajectory at $\alpha_{0'}$. It then merges onto the Regge trajectories at $\alpha_{1}$ given in Eq.~\eqref{eq:alpha_1}. Then it follows this trajectory until $M=M_\mathrm{de}^{(011)}$. 
%%%
	\item $\alpha_{0'} < \alpha_{0}$: 
	This case corresponds to the scenario in Sec.~\ref{sec:acc_g_GW} and the curves with $f_\text{Edd}=10^{-9}$ in Fig.~\ref{fig:Regge}.	The system is dominated by superradiance initially and quickly merges to the Regge trajectories at $M = M_1$. Here $M_1$ is estimated using Eqs.~\eqref{eq:Ms011_max_WoAcc} and \eqref{eq:conservation_M}. The system then follows the $\{0,1,1\}$ Regge trajectory until $M=M_\mathrm{de}^{(011)}$.  
\end{itemize}
%%%%%%

Finally, we turn to what is not reflected in the Regge plots. The BH mass grows almost exponentially before merging onto the Regge trajectory, from which one could obtain the corresponding time cost. This can then be applied to calculate the condensate mass and GW emission rate. When the BH evolves along the Regge trajectory, the Eqs.~\eqref{eq:Edot_SR} to \eqref{eq:t_3} can be directly applied for estimating the important quantities in this phase.

With these equations, one could understand why the condensate mass evolution and GW emission flux are dramatically affected when $\tau_\text{acc}\lesssim \tau_\text{SR}$, which is shown in Fig.~\ref{fig:vs_fEdd}. Without accretion, the BH mass $M$ is roughly constant, resulting in an exponential increase of the condensate mass. The time duration of the GW is controlled by $\tau_\text{GW}$. Comparably, the condensate mass with accretion is a double exponential function and the termination of the GW emission is controlled by $\tau_\text{de}\ll \tau_\text{GW}$. The maximum condensate mass is also larger with accretion, which reaches as much as 0.15 in Fig.~\ref{fig:single}, while this number never exceeds 0.1 if accretion is absent. Consequently, the GW emission in Eq.~\eqref{eq:dot_E_GW} is stronger in magnitude but shorter in time.

\section{Multi-mode evolution with accretion}\label{sec:multi}

In this section, we study the evolution in the presence of multiple modes. The evolution can be separated into stages with different $l$. In each $l$-stage, the modes with the same azimuthal number $l$ grow the fastest via superradiance. We keep only the dominant ($\{0,l,l\}$) and the subdominant ($\{1,l,l\}$) modes in each $l$-stage for illustration, while the generalization to more modes is straightforward. The superradiance rates of modes with azimuthal numbers larger than $l$ are orders of magnitude smaller and thus can be safely ignored in the $l$-stage. From the experience learned in the previous section, we start with the BH evolution on the Regge plot in Sec.~\ref{sec:Regge_Multi}. Readers not interested in technical details can only read this subsection. Analytical approximations for the important quantities are given in Sec.~\ref{sec:Multi_BH_Dom}. Then the subdominant modes and GW emission flux are discussed in Sec.~\ref{sec:Multi_sub} and Sec.~\ref{sec:Multi_GW}, respectively.

\subsection{Evolution on the Regge plot}\label{sec:Regge_Multi}

%%%%%FFFFF
\begin{figure*}[htbp]
	\centering 
	\includegraphics[width=1\textwidth]{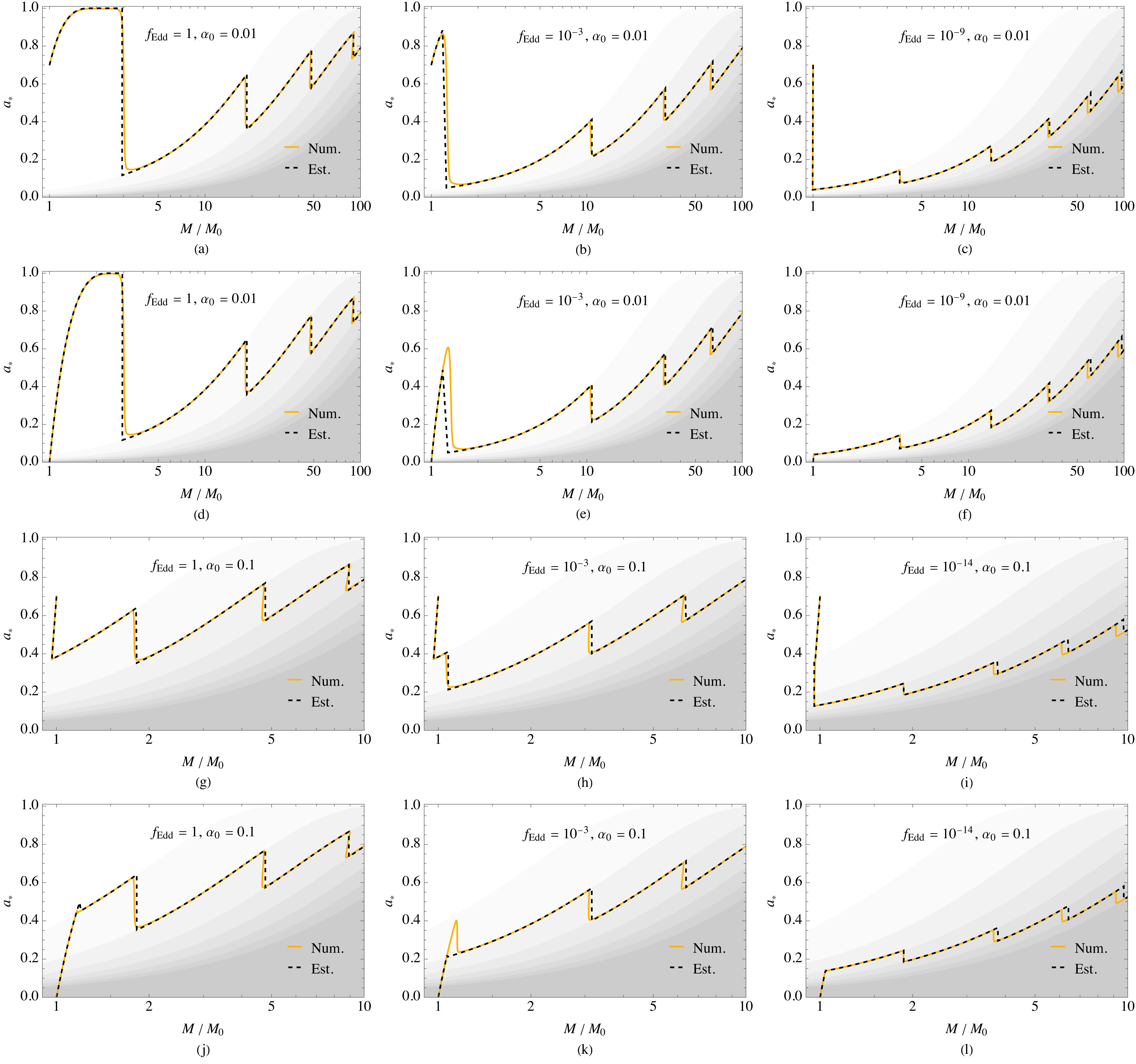}
	\caption{The dependence of BH spins on the BH masses for different initial parameters and accretion rates. The initial condensate masses are $\M[(nll)]{M}{s,0}=10^{-5}M_0$ with $n=0,1$ and $1\leq l\leq 7$. The accretion rate and the initial mass coupling are shown in the legends of each panel. The initial BH mass and mass coupling for panels (a-f) are $M_0=10^3M_\odot$ and $\alpha_0=0.01$, respectively. For panels (g-l), the initial BH mass is $10^4 M_\odot$ and the initial mass coupling is $\alpha_0=0.1$. The initial BH spin is $a_{*0}=0.7$ for panels (a-c,g-i) and $a_{*0}=0$ for panels (d-f,j-l). The upper boundaries of the shaded areas from the top to bottom are the Regge trajectories of the $\{0,l,l\}$ modes with $l$ increasing from 1 to 7.} 
	\label{fig:Regge_Multi} 
\end{figure*}
%%%%%

The multi-mode evolution with different initial conditions and accretion rates are shown in Fig.~\ref{fig:Regge_Multi}. Compared to the single mode case in Fig.~\ref{fig:Regge}, the BH evolution with multiple modes is a little more complicated but still has a quite simple pattern. The BH may follow the accretion trajectory for a while then transits quickly to a Regge trajectory. The time spent on the accretion trajectory depends on the initial BH mass and spin, as well as the accretion rate. If the accretion rate is small, the BH spin may drop since $t=0$. Nonetheless, with multiple modes, the BH does not follow a single Regge trajectory all the way to the critical mass. Instead, it steps down to the Regge curves with an increasing value of $l$. To have a good understanding of the evolution, we need to have satisfactory estimates of the critical points where the BH leaves or merges into a trajectory. These critical points split the whole evolution into different time segments. In each segment, the unimportant effects could be ignored and the picture is much simpler to analyse.

It is important to figure out which Regge trajectories may be activated in the evolution, i.e., the corresponding modes grow via superradiance for some time. This is not a simple task and requires several steps in the presence of multiple modes. First of all, it is simple to identify those Regge trajectories for which the superradiance condition $\omega_{0ll}<l\Omega_H$ are never met. This gives a critical value of $l$, referred as $l_\text{c}$ below, defined as the minimum integer satisfying
%%%%%=====
\begin{align}\label{eq:lc}
	M_0\omega_{0ll} < l/2.
\end{align}
%%%%%
The modes with $l<l_\text{c}$ are in their decaying phase since $t=0$ and do not affect the evolution of the BH-condensate system. 

We first analyze the case when the initial BH spin $a_{*0}$ is greater than the $a_{*\text{c}}^{(0l_\mathrm{c}l_\mathrm{c})}$ given in Eq.~\eqref{eq:a_critc}. If the accretion rate is much larger than the superradiance rate of the $l_c$ mode, the BH first follows the accretion trajectory to $\sim a_{*\text{lim}}$ before transiting to the Regge curve with $l=l_\text{c}$, see Fig.~\ref{fig:Regge_Multi}(a). With a decreasing value of the accretion rate, the time cost on the accretion trajectory is shorter, see Fig.~\ref{fig:Regge_Multi}(b). At some point, the BH spin converges to the $l_\text{c}$ Regge curve since $t=0$, see Fig.~\ref{fig:Regge_Multi}(c). Further decreasing the accretion rate and/or increasing the superradiance rate, the time cost along the $l_c$ Regge trajectory is shorter, see Fig.~\ref{fig:Regge_Multi}(g,h). Until at some point, the accretion effect is negligible during the time when the BH is on the $l_\text{c}$ Regge curve. Then the BH does not evolve along the $l_\text{c}$ Regge curve before stepping down to the $l_\text{c}+1$ Regge trajectory, see Fig.~\ref{fig:Regge_Multi}(i).

Given the initial conditions, we need to judge which pattern the later evolution has. To separate these cases, one first calculates $\alpha_{0'}^{(0l_\text{c}l_\text{c})}$, which is proportional to the BH mass at the point when its spin reaches the local maximum close to the trajectory curve. The formula for $l_\text{c}=1$ is given in Eq.~\eqref{eq:alpha_0p}, and those for larger values of $l_\text{c}$ are listed in Eqs.~\eqref{eq:alpha_0p_0ll}. When $\mu\tau_\text{acc}$ is large, the asymptotic expression of the Lambert $W_0$ function could be useful
%===
\begin{align}\label{eq:W0_asym}
\lim_{x\to +\infty} W_0(x) = \log x - \log\left(\log x\right).
\end{align}
%===
When $\mu\tau_\text{acc}$ is small, the $\alpha_{0'}$ is significantly larger than $\alpha_0$. With an increasing value of $\mu\tau_\text{acc}$, it first decreases to below $\alpha_0$, then approaches $\alpha_0$ from below. So there are three possibilities
%%%%%%%
\begin{itemize}
	\item $\alpha_{0'}\geq\sqrt{6}\,k\alpha_0$: 
	The BH spin can reach the plateau region as in Fig.~\ref{fig:Regge_Multi}(a).  
%%%
	\item $\alpha_0<\alpha_{0'}< \sqrt{6}\,k\alpha_0$: 
The BH spin follows the accretion trajectory for a while and then drops before reaching $a_{*\text{lim}}$, as in Fig.~\ref{fig:Regge_Multi}(b).
%%%
	\item $\alpha_{0'}<\alpha_0$:
	The BH spin decreases since $t=0$ as in Fig.~\ref{fig:Regge_Multi}(c,g-i).
\end{itemize}
%%%%%%%

In any of the above three cases, the evolution merges onto the $\{0,l_\text{c}, l_\text{c}\}$ Regge curve. With only a single mode as explained in Sec.~\ref{sec:single}, the BH spin reaches a local minimum close to the Regge curve. There the time of the local minimum is defined as $t_1$, and we also define the quantities at this time using the subscript, such as $\alpha_1$. In the presence of multiple modes, there is a local minimum of the BH spin at each Regge curve. We similarly define this time at the $\{0,l,l\}$ Regge curve as $t_1^{(0ll)}$ and the corresponding quantities using an additional superscript $(0ll)$, such as $\alpha_1^{(0ll)}$. To avoid confusion, the mass of a superradiant mode at a certain time is written explicitly. For example $M_\text{s}^{(022)}(t_1^{(011)})$ denotes the mass of the $\{0,2,2\}$ mode at time $t_1^{(011)}$. 

The value of $\alpha_1^{(0l_\text{c}l_\text{c})}$ can be obtained using Eqs.~\eqref{eq:alpha_1_0lclc}. One could then calculate the value of $\alpha_1^{(0ll)}$ with $l>l_\text{c}$ using Eqs.~\eqref{eq:alpha_1_0ll}. Then the evolution on the Regge plot can be approximately fixed with the values of $\alpha_{0'}$ and $\alpha_1^{(0ll)}$.
When $\mu\tau_\text{acc}$ is large, the following asymptotic expression for the Lambert $W_{-1}$ function is useful
%===
\begin{align}
\lim_{x\to 0^-} W_{-1}(x) = \log(-x) - \log\left(-\log(-x)\right).
\end{align}
%===

In the limit of $\mu\tau_\text{acc}\to +\infty$, the obtained $\alpha_1^{(0,l+1,l+1)}$ approaches $\alpha_1^{(0ll)}$ from above. Then the evolution does not follow the $\{0,l,l\}$ Regge curve. This may happen at the first several Regge curves, causing an approximate vertical drop of the evolution curve on the Regge plot since $t=0$. In this time range, a better treatment is to ignore the accretion compared to the superradiance. One could then use the energy and angular momentum conservation for the transition between two consecutive Regge curves
%%%%%=====
\begin{subequations}\label{eq:conservation_l}
	\begin{align}
		\M[(0,l+1,l+1)]{a}{*c} \left(\alpha^{(0,l+1,l+1)}_1\right)^2 + (l+1)\mu \M[(0,l+1,l+1)]{M}{s}(t_1^{(0ll)}) &= \M[(0ll)]{a}{*c} \left(\alpha^{(0ll)}_1\right)^2 + l \mu \M[(0ll)]{M}{s}(t_1^{(0ll)}),
	\\
		\alpha_1^{(0,l+1,l+1)} + \mu\M[(0,1+1,1+1)]{M}{s}(t_1^{(0ll)}) &= \alpha_1^{(0ll)} + \mu\M[(0ll)]{M}{s}(t_1^{(0ll)}).
	\end{align}
\end{subequations}
%%%%%
The $\alpha_1^{(0ll)}$ obtained from these conservation equations agrees better with the numerical results, which can be observed in Fig.~\ref{fig:Regge_Multi}(c,g-i). Since the conservations give $\alpha_1^{(0,l+1,l+1)}\approx 0.9\, \alpha_1^{(0ll)}$ \cite{Guo:2022mpr}, we claim Eqs.~\eqref{eq:conservation_l} give a more accurate description when $\alpha_1^{(0,l+1,l+1)}-\alpha_1^{(0ll)}\ll 0.1 \,\alpha_1^{(0ll)}$.

If $a_{*0}<a_{*\text{c}}^{(0l_\mathrm{c}l_\mathrm{c})}$, the situation is a little more complicated. One first find the value of $l'>l_\text{c}$ which satisfies 
%===
\begin{align}\label{eq:l_prime}
a_{*\text{c}}^{(0,l'+1,l'+1)}<a_{*0}<a_{*\text{c}}^{(0,l',l')}.
\end{align}
%===
Then the modes with $l\in [l_\text{c},l']$ do not satisfy the superradiance condition initially. These modes shrink in size at first, transferring the energy and angular momentum to the BH. If accretion is strong, the BH spin could climb along the accretion trajectory to above the $\{0,l_\text{c},l_\text{c}\}$ Regge curve, see Fig.~\ref{fig:Regge_Multi}(d,e). With a smaller accretion rate and/or larger superradiance rate, the BH spin can just reach the $\{0,l_\text{c},l_\text{c}\}$ Regge curve, see Fig.~\ref{fig:Regge_Multi} (f,j). If we further decrease the accretion rate compared to the superradiance rate, the BH spin can only reach some Regge curve with $l>l_\text{c}$, see Fig.~\ref{fig:Regge_Multi} (k,l).

We define $l_\text{acc}$ as the first Regge curve which the evolution follows along. Its value could be larger, smaller, or equal to $l'$, depending on the relative size of the accretion rate and the superradiance rates of different modes. We first judge whether $l_\text{acc}$ is larger than $l'$. After many tests, the best scheme we could find to calculate $l_\text{acc}$ is by comparing the $\alpha_{0'}^{(0ll)}$ with different $l$. Here $\alpha_{0'}^{(0ll)}$ is the mass coupling at the time when the BH spin reaches its local maximum if only accretion and the $\{0,l,l\}$ mode are present, which is given by Eqs.~\eqref{eq:alpha_0p_0ll}. Specifically, one calculate the $\alpha_{0'}^{(0ll)}$ with increasing value of $l$ from $l_\text{c}$ to $l'$. If a minimum of $\alpha_{0'}^{(0ll)}$ is observed for $l<l'$, then the BH spin reaches the local maximum between the $\{0,l-1,l-1\}$ and $\{0,l,l\}$ Regge curves. In this case, we set $l_\text{acc}=l$. After identifying $l_\text{acc}$, one could reset $t=0$ as the time when the BH reaches the intersection of the accretion trajectory and the $\{0,l_\text{acc},l_\text{acc}\}$ Regge curve. Then the later evolution can be described in the same way as the $a_{*0}>a_{*\text{c}}^{(0l_\text{c}l_\text{c})}$ case with $l_\text{c}$ replaced by $l_\text{acc}$.

On the other hand, if the minimum value of $\alpha_{0'}^{(0ll)}$ is observed at $l=l'$, one could ignore all the Regge curves with $l<l'$. Then the evolution is the same as the $a_{*0}>a_{*\text{c}}^{(0l_\text{c}l_\text{c})}$ case with $l_\text{c}$ replaced by $l'$.

The evolutions with $a_{*0}=0$ are presented in \fref{fig:Regge_Multi}, panels (d-f) and panels (j-l), where other initial parameters are the same as panels (a-c) and panels (g-i), respectively. In this case, the $l'$ is infinity. Analytical estimates of the points where the evolution leaves a trajectory and merges into the next are given in Sec.~\ref{sec:Multi_BH_Dom}. For various initial conditions and accretion rates, the analytical approximations agree quite well with the numerical results.

\subsection{Evolution of the BH and dominate modes}
\label{sec:Multi_BH_Dom}

The BH evolution is mainly determined by the dominant mode and/or the accretion in each $l$-stage. In this part, we focus on the dominant modes, while leaving the discussion about the subdominant modes in Sec.~\ref{sec:Multi_sub}. To help the analysis, we present the numerical results for a specific multi-mode scenario in \fref{fig:Multi}. The initial parameters are $M_0=10^4 M_\odot$, $\alpha_0=0.1$, $M_\text{s,0}^{(nll)}=10^{-5}M_0$ and $f_\text{Edd}=10^{-3}$. The value of $f_\text{Edd}$ is chosen such that the accretion time scale ${\tau}_\text{acc}$ is at the same order of $\tau_\text{SR}^{(022)}$, which is the the superradiance time scale of $(0,2,2)$ mode. This time scale is much larger than $\tau_\text{SR}^{(011)}$ but much smaller than $\tau_\text{SR}^{(033)}$.

Two initial BH spins  $a_{*0}=0.7$ and $a_{*0}=0$ are considered. From Eq.~\eqref{eq:lc}, it is easy to see $l_\text{c}=1$. Then $a_{*\text{c}}^{(011)}\approx 0.384$ is calculated from Eq.~\eqref{eq:a_critc} using the value of $\alpha_0$.

\subsubsection{\texorpdfstring{$a_{*0}=0.7$}{a0>ac}}
\label{sec:Multi_a0_g_ac}

This case corresponds to $a_{*0}>a_{*\text{c}}^{(0l_\mathrm{c}l_\mathrm{c})}$. We first identify the evolution of the BH on the Regge plot. From Eq.~\eqref{eq:tau_acc}, one could obtain $\tau_\text{acc}\mu=3.2\times 10^{18}$. We then apply Eq.~\eqref{eq:alpha_0p} and get $\alpha_{0'}=\alpha_0-8\times 10^{-10}$, which is less than $\alpha_0$. Thus the BH does not evolve along the accretion trajectory at all, corresponding to the third scenario in the discussion below Eq.~\eqref{eq:W0_asym}.

The $\alpha_1^{(011)}$ can be calculated using Eq.~\eqref{eq:alpha_1}. Since the BH does not evolve along the accretion trajectory, the $M_{\text{s},0''}$ $a_{*0''}$, and $\alpha_{0''}$ should be replaced by $M_\text{s,0}$, $a_{*0}$, and $\alpha_0$, respectively. Using the above parameters, one obtains $\alpha_1^{(011)}=\alpha_0+7\times 10^{-10}$. The small values of $\alpha_{0'}-\alpha_0$ and $\alpha_1^{(011)}-\alpha_0$ indicate the evolution merges onto the $\{0,1,1\}$ Regge curve almost at $t=0$.

We continue to calculate $\alpha_1^{(022)}$. The steps are quite similar to those from Eq.~\eqref{eq:dasdt} to Eq.~\eqref{eq:alpha_1}. The time derivative of the BH spin in the presence of multiple modes is
%===
\begin{align}\label{eq:dasdt_multi}
	\dot{a}_* &= \frac{1}{\tau_\text{acc}}\left(\frac{J_\mathrm{ms}^{\dagger}}{M E_\mathrm{ms}^{\dagger}}-2a_*\right)
	-\sum_{l=l_\text{c}}^\infty \frac{2M_\text{s}^{(0ll)}}{M}\left(\frac{l}{M\omega_{0ll}}-2a_*\right)\Gamma_{0ll},
\end{align}
%===
where we have only kept the $\{0,l,l\}$ modes. Other terms have much smaller effect on the BH evolution and could be safely ignored. In the time range from $t_1^{(011)}$ to $t_1^{(022)}$, only the accretion and the $l=1,2$ modes are important. When the BH follows the $\{0,1,1\}$ Regge curve, the accretion almost balances the superradiance of the $\{0,1,1\}$ mode,  resulting in a rather slow change of $a_*$. During this time, the $\{0,2,2\}$ mode is unimportant due to its small mass. When the $\{0,2,2\}$ mode is dense enough, it dominates the evolution, causing the rapid drop of $a_*$ to the $\{0,2,2\}$ curve close to $t_1^{(022)}$. Thus as a LO estimate, one could only consider the contribution of the $\{0,2,2\}$ mode on the right side of Eq.~\eqref{eq:dasdt_multi}
%===
\begin{align}\label{eq:dasdt_022}
\dot{a}_* = -\frac{2M_\text{s}^{(022)}}{M^2\mu}\Gamma_{022}.
\end{align}
%===
One could take the LO of  $\Gamma_{022}$ in $\alpha$ expansion with $a_*=a_{*\text{c}}^{(011)}$, which is
%%%%%=====
\begin{align}\label{eq:Gamma022}
	M\Gamma_{022} = \frac{64 \alpha^{14}}{885735}.
\end{align}
%%%%%
The $M_\text{s}^{(022)}$ can be calculated by inserting Eqs.~\eqref{eq:Gamma022} and \eqref{eq:BH_mass_efold} into Eq.~\eqref{eq:Ms_ODE} with the GW radiation ignored
%===
\begin{align}
\frac{M_\text{s}^{(022)}}{M_\text{s}^{(022)}(t_1^{(011)})}
=\exp\left\{\frac{128 \M[]{\tau}{acc} \mu }{11514555} \left[\alpha^{13}-\left(\alpha_{1}^{(011)}\right)^{13}\right]\right\},
\end{align}
%===
which is valid for $t\in[t_1^{(011)}, t_1^{(022)}]$. For the current example, there is $M_\text{s}^{(022)}(t_1^{(011)}) \approx M_\text{s,0}^{(022)}$. Then Eq.~\eqref{eq:dasdt_022} can be integrated directly from $t_1^{(011)}$ to $t_1^{(022)}$. Inserting the values of $a_{*\text{c}}^{(0ll)}$ in Eq.~\eqref{eq:a_critc}, one gets 
%%%%%=====
\begin{align}\label{eq:alpha_1_022}
	\alpha_1^{(022)} \approx \left\{-\frac{5\cdot3^{12}}{2^{7}\tau_\mathrm{acc}\mu}   W_{-1}\left[-\frac{128\tau_{\mathrm{acc}}\mu}{2657205}\left(M_{\mathrm{s0}}^{(022)}\mu e^{-\frac{128\left(\alpha_{1}^{(011)}\right)^{13}\tau_{\mathrm{acc}}\mu}{11514555}}\right)^{\frac{13}{3}}\right]\right\}^{\frac{1}{13}},
\end{align}
%%%%%
where the term with the exponential integral function is ignored for the same reason discussed below Eq.~\eqref{eq:alpha_1}.

The $\alpha_1^{(0ll)}$ for larger $l$ can be calculated similarly, which are listed in Eqs.~\eqref{eq:alpha_1_0ll}. For the current example, after inserting the numbers, one arrives at $\alpha_1^{(022)}=0.109$, $\alpha_1^{(033)}=0.315$ and $\alpha_1^{(044)}=0.639$.

Since $\alpha_1^{(011)}-\alpha_0\ll 0.1 \alpha_0$, one could ignore accretion and apply the energy and angular momentum conservations in Eq.~\eqref{eq:conservation_l} for more accurate estimates. Inserting the values, one arrives at $\alpha_1^{(011)}=0.097$. The value of $\alpha_1^{(022)}$ in Eq.~\eqref{eq:alpha_1_022} is adjusted accordingly to $0.108$, which satisfies $\alpha_1^{(022)}-\alpha_1^{(011)}>0.1 \alpha_1^{(011)}$. Then $\alpha_1^{(0ll)}$ for $l\geq 3$ could be calculated using Eqs.~\eqref{eq:alpha_1_0ll}. Connecting these values on the accretion trajectory as well as the Regge curves, one could get an approximate evolution on the Regge plot, which is shown in Fig.~\ref{fig:Regge_Multi}(h).

We now discuss the evolution of the dominant modes. Since it is the same for all $l$-stage, the presentation below works for any $\{0,l,l\}$ mode with $l\geq l_\text{c}$. Between $t_1^{(0ll)}$ and $t_1^{(0,l+1,l+1)}$, the $\{0,l,l\}$ mode first grows, reaching the maximum mass ratio $M_\text{s}^{(0ll)}/M$ at $t_2^{(0ll)}$, then gradually shrinks due to GW emission. Finally, it enters into the decaying phase and quickly falls into the BH close to $t_1^{(0,l+1,l+1)}$. In this time range, one could only consider the $\{0,l,l\}$ and $\{0,l+1,l+1\}$ modes. Adding the modes with the same $l$ but $n > 0$ has little effect on the BH evolution. These modes, especially the subdominant modes with $n=1$, are important in the GW waveform, which will be discussed in the next subsection.

The analysis is quite similar to that from Eq.~\eqref{eq:Edot_SR} to \eref{eq:t_2}. Between $t_1^{(0ll)}$ and $t_1^{(0,l+1,l+1)}$, the BH evolves along the $\{0,l,l\}$ Regge curve. For most of this time range, the evolution is dominated by the accretion and the $\{0,l,l\}$ mode in Eq.~\eqref{eq:dasdt_multi}, with other terms safely neglected. On the other hand, the $\dot{a}_*$ can also be calculated using the  Eq.~\eqref{eq:a_critc}. Combining these two expressions of $\dot{a}_*$, one could solve $\dot{E}_\text{SR}^{(0ll)}$ as a function of $\alpha$. Then taking GW emission in Eq.~\eqref{eq:Ms_ODE} as a perturbation and assuming $\alpha$ increases with time as in Eq.~\eqref{eq:BH_mass_efold}, one could further solve $M_\text{s,Regge}^{(0ll)}$ and the mass of the $\{0,l,l\}$ mode in this time range 
%===
\begin{align}\label{eq:Ms_0ll}
	M_\text{s}^{(0ll)}(t)
	=M_\text{s}^{(0ll)}(t_1^{(0ll)})
	+M_\text{s,Regge}^{(0ll)}(\alpha)
	-M_\text{s,Regge}^{(0ll)}(\alpha_1^{(0ll)})
	+\delta M_\text{s}^{(0ll)}(t).
\end{align}
%===
The expressions of $M_\text{s,Regge}^{(0ll)}$ and $\delta M_\text{s}$ for $l\leq 7$ are given in Eqs.~\eqref{eq:Ms_Regge_0ll} and \eqref{eq:delta-Ms_Regge_0ll}, respectively.

The value of $\alpha_2^{(0ll)}$ can be determined with $d\left(M_\text{s}^{(0ll)}/M\right)/dt=0$ at $t_2^{(0ll)}$, which gives an equation like Eq.~\eqref{eq:for_alpha_2}, with the superscript $(011)$ replaced by $(0ll)$. Then one could solve $\alpha_2^{(0ll)}$ from this equation. Using the value of $\alpha_2^{(0ll)}$, one could estimate $t_2^{(0ll)}$ by assuming the BH mass grows as in Eq.~\eqref{eq:BH_mass_efold} from $t=0$ to $t_2^{(0ll)}$.
A more accurate estimate can be obtained by inserting the $\dot{E}_\text{Regge}^{(0ll)}$ into Eq.~\eqref{eq:M_BH_ODE} and solving the differential equation, in the same manner as Eqs.~\eqref{eq:dot_M_corr} and \eqref{eq:M_g_t_1}
%%%%%=====
\begin{align}\label{eq:M_g_t_1_0ll}
	M  \approx \frac{2 l \alpha_1^{(0ll)} e^{(t-t_1^{(0ll)})/\tau  }/ \mu}{2l + 3 \sqrt{6} \alpha_1^{(0ll)} \left[e^{(t-t_1^{(0ll)})/\tau } - 1\right]}.
\end{align}
%%%%%
Finally, one can solve out $t_2^{(0ll)}$ with this expression of $M(t)$ and the value of $\alpha_2^{(0ll)}$. The obtained $\alpha_2^{(0ll)}$ are listed in Eqs.~\eqref{eq:alpha_2_0ll} for $l\leq 7$. Inserting these values into Eq.~\eqref{eq:Ms_0ll}, the maximum mass ratio of the $\{0,l,l\}$ mode could be calculated.

Combining the obtained expression of $M(t)$ with the value of $\alpha_1^{(0ll)}$, one could obtain a better estimate of 
%===
\begin{align}
\Delta t^{(0ll)}\equiv t^{(0,l+1,l+1)}_1-t^{(0ll)}_1,
\end{align}
%===
which is the time duration of the $l$-stage and is used to estimate the duration of the GW emission of the $\{0,l,l\}$ mode below.

For our example, the above approximations are compared to the numerical results in Table.~\ref{tab:Multi}. For all the quantities, these approximations work reasonably well.

\subsubsection{\texorpdfstring{$a_{*0}=0$}{a0<ac}}
\label{sec:Multi_a0_l_ac}

In the case of $a_{*0}<a_{*\text{c}}^{(0,l_\text{c}l_\text{c})}$, one first identify the value of $l'$ with Eq.~\eqref{eq:l_prime}, which is $\infty$ for $a_{*0}=0$.

To get $l_\text{acc}$, we calculate $\alpha_{0'}^{(0ll)}$ with $l$ increasing from $l_\text{c}$ to $l'$. Here $\alpha_{0'}^{(0ll)}$ is the mass coupling at the time when the BH spin reaches the local maximum before transiting to the $\{0,l,l\}$ Regge curve if only accretion and the $\{0,l,l\}$ mode are present. For $l<l'$, the calculation of $\alpha_{0'}^{(011)}$ starts from the intersection of the accretion trajectory and the $\{0,l,l\}$ Regge curve, which is given in Eq.~\eqref{eq:intersection}. We start from $l=m=1$, the mass coupling at the intersection is $\alpha_\text{int}^{(011)}=0.116$. The mass of the cloud can be estimated using Eq.~\eqref{eq:Ms011_x_l_sqrt6}. For our example, one gets $M_\text{s,int}^{(011)} = 6\times 10^{-25987594}\approx 0$. Then taking the intersection as the new initial time, one could eventually apply Eq.~\eqref{eq:alpha_0p} and get $\alpha_{0'}^{(011)}=0.117$.

Similarlly, one could calculate the $M_\text{s}^{(0ll)}$ and $\alpha_{0'}^{(0ll)}$ for any $\{0,l,l\}$ modes. Their expressions for $l\leq 7$ are listed in Eqs.~\eqref{eq:Ms_accretion_trajectory} and~\eqref{eq:alpha_0p_0ll}. For our example, we have $\alpha_{0'}^{(022)}=0.107$ and $\alpha_{0'}^{(033)}=0.232$. Therefore, we conclude $l_\text{acc}=2$.

Then one could choose the intersection of the accretion trajectory and the $\{0,2,2\}$ Regge curve as $t=0$, all the steps explained in Sec.~\ref{sec:Multi_a0_g_ac} can be applied for the later evolution. The comparison of the analytical approximation with the numerical results is shown in Fig.~\ref{fig:Regge_Multi}(k).

%%%%%FFFFF
\begin{figure*}[htbp]
	\centering 
	\includegraphics[width=0.9\textwidth]{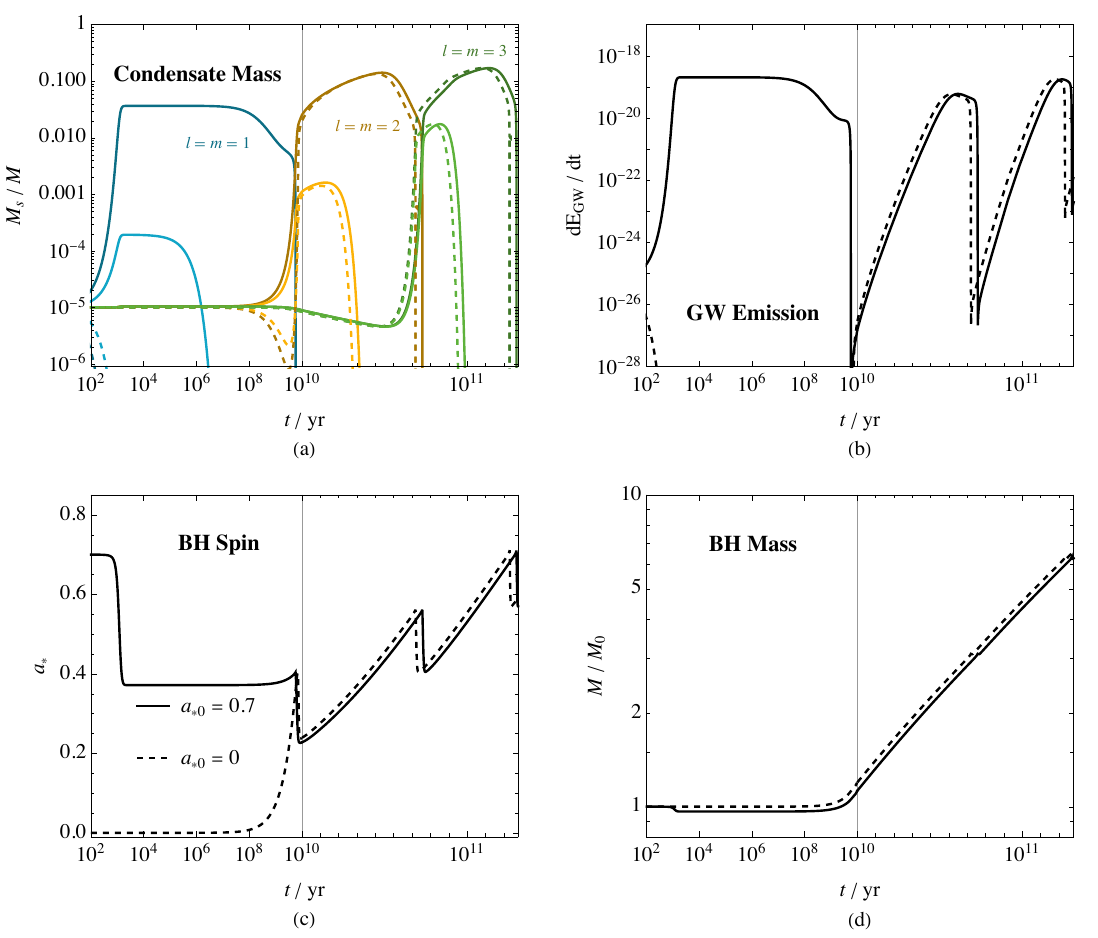}
	\caption{The time evolution of scalar condensate masses (panel a), total GW emission fluxes (panel b), BH spins (panel c), and BH masses (panel d). The initial parameters are $M_0=10^4 M_\odot$, $\M[(nll)]{M}{s,0}=10^{-5}M_0$, $\alpha_0=0.1$, and $\M[]{f}{Edd}=10^{-3}$. The solid and dashed curves are for initial BH spins $a_{*0}=0.7$ and $a_{*0}=0$, respectively. In panel (a), the darker curves represent the dominant modes, while the lighter curves represent the subdominant modes. For better presentation, the horizontal axis employs a linear scale for $t<10^{10}$~yr and a logarithmic scale for $t>10^{10}$~yr.} 
	\label{fig:Multi} 
\end{figure*}
%%%%%

%%%%%TTTTT
\begin{table*}[htbp]\footnotesize
	\centering
	\renewcommand\arraystretch{1.2}
	\begin{tabular}{p{0.6cm}<{\centering}p{0.6cm}<{\centering}p{1.8cm}<{\centering}p{1.7cm}<{\centering}p{1.7cm}<{\centering}p{1.7cm}<{\centering}p{1.8cm}<{\centering}p{1.7cm}<{\centering}}
	\hline
	\hline
 	\multicolumn{2}{c}{$l$} & $(\M[(0ll)]{M}{s}/M)_\mathrm{max}$ & $\alpha_{2}^{(0ll)}$ & $\M[(0ll)]{\tau}{GW}$ / yr & $\M[(1ll)]{M}{s,max}/\M[(0ll)]{M}{s}$ & $\alpha_{2'}^{1ll}$ or $\alpha_{1}^{(0ll)}$ & $\M[(1ll)]{\tau}{life}$ / yr \\
	\hline
	\multirow{2}{*}{$l=1$} & Est. & $3.67\times10^{-2}$ & $9.65\times10^{-2}$ & $2.84\times10^{8}$ & $4.98\times10^{-3}$ & $9.65\times10^{-2}$ & $5.71\times10^{5}$ \\
	& Num. & $3.66\times10^{-2}$ & $9.65\times10^{-2}$ & $2.92\times10^{8}$ & $5.28\times10^{-3}$ & $9.65\times10^{-2}$ & $1.08\times10^{6}$\\
	\hline
	\multirow{2}{*}{$l=2$} & Est. & $1.33\times10^{-1}$ & $2.18\times10^{-1}$ & $8.60\times10^{10}$ & $2.70\times10^{-2}$ & $1.41\times10^{-1}$ & $2.93\times10^{10}$ \\
	& Num. & $1.41\times10^{-1}$ & $2.24\times10^{-1}$ & $7.13\times10^{10}$ & $2.36\times10^{-2}$ & $1.43\times10^{-1}$ & $2.57\times10^{10}$\\
	\hline
	\multirow{2}{*}{$l=3$} & Est. & $1.46\times10^{-1}$ & $4.93\times10^{-1}$ & $8.71\times10^{10}$ & $2.52\times10^{-1}$ & $3.46\times10^{-1}$ & $1.95\times10^{10}$ \\
	& Num. & $1.70\times10^{-1}$ & $5.11\times10^{-1}$ & $5.01\times10^{10}$ & $2.86\times10^{-1}$ & $3.54\times10^{-1}$ & $1.52\times10^{10}$\\
	\hline
	\end{tabular}
	\caption{Comparison of the analytical estimates to the numerical results obtained from solving \eqref{eq:total_ODEs} for essential quantities. Initial parameters are identical to those used for the solid lines in Fig.~\ref{fig:Multi}.}
	\label{tab:Multi}
\end{table*}
%%%%%%%%%%

\subsection{Subdominant modes}
\label{sec:Multi_sub}

The subdominant modes with indices $\{1,l,l\}$ are important for GW beat signature. In this part, we explain the calculation of the maximum mass and the lifetime of these modes. The critical BH spin of the subdominant mode $a_*^{(1ll)}$ can be directly calculated with Eq.~\eqref{eq:a_critc}. It is a little larger than that of the dominant mode with the same value of $l$. As a result, the $\{1,l,l\}$ Regge curve is a bit above the $\{0,l,l\}$ one. Then at the beginning of the $l$-stage when the BH spin quickly decreases, the evolution first reaches the $\{1,l,l\}$ curve. It could follow along the $\{1,l,l\}$ curve for a while before stepping down to the $\{0,l,l\}$ curve. It is also likely that the evolution quickly crosses the $\{1,l,l\}$ curve and merges to the $\{0,l,l\}$ curve. In both cases, the growth time of the subdominant mode is shorter than the dominant one. In this work, we assume the initial mass of the subdominant mode is not unnaturally large. Then one could conclude that the mass of the subdominant mode is always much smaller than the corresponding dominant mode for $l\leq 2$ since the superradiance rate of the former is always smaller than the latter. For $l\geq 3$, there exists a ``overtone mixing" phenomenon where the superradiance rate of the subdominant mode is comparable to, or even larger than, the dominant mode  \cite{Siemonsen:2019ebd}. Interestingly, the above analytical approximations still work for the correct order of magnitude. We will discuss this scenario in Sec.\ref{sec:summary}. In the rest of this section, we assume the subdominant mode always has a smaller mass than the corresponding dominant mode.

Because of the small mass, the subdominant modes do not play an important role in the BH evolution, qualifying the above analysis with only the dominant modes. Nonetheless, the subdominant modes are important for the GW beat signature. Below we take $l=1$ stage as an example. Stages with other values of $l$ could be analyzed in the same way and we simply present the results in the appendix.

We first study the maximum mass of the subdominant mode. To be consistent in the notation with the derivation of the dominant mode, we define this time as $t_2^{(111)}$. Since it must be smaller than $t_2^{(011)}$, the GW emission can be safely ignored before $t_2^{(111)}$. We first assume it is larger than $t_1^{(011)}$ when the BH spin reaches the local minimum. If both $\{0,1,1\}$ and $\{1,1,1\}$ modes are included, the left side of Eq.~\eqref{eq:Edot_SR} should be changed to $\dot{E}_\text{SR}^{(011)}+\dot{E}_\text{SR}^{(111)}$. At time $t_2^{(111)}$, the BH spin equals to $a_{*\text{c}}^{(111)}$ given in Eq.~\eqref{eq:a_critc}, causing $\dot{E}_\text{SR}^{(111)}=0$. On the other hand, $\dot{E}_\text{SR}^{(011)}=2M_\text{s}^{(011)}\Gamma_{011}$, in which $\Gamma_{011}$ is given in Eq.~\eqref{eq:Gamma011} and $M_\text{s}^{(011)}$ can be estimated by
%===
\begin{align}
	M_\text{s}^{(011)}(\alpha_2^{(111)}) \approx & \M[(011)]{M}{s,1}+ \M[(011)]{M}{s,Regge}(\alpha_2^{(111)})- \M[(011)]{M}{s,Regge}(\alpha_1^{(011)}),
\end{align}
%===
with $\M[(011)]{M}{s,Regge}$ given in Eq.~\eqref{eq:Ms011_Regge}. Then one could solve for the value of $\alpha_2^{(111)}$ at the LO of $\alpha_0$
%===
\begin{align}
\alpha^{(111)}_{2} &\approx \left(\frac{864}{5 \mu \M[]{\tau}{acc} }\right)^{1/11}.
\end{align}
%===

If the obtained $\alpha_2^{(111)}$ is greater than $\alpha_1^{(011)}$, it is consistent with our assumption $t_2^{(111)}>t_1^{(011)}$. It indicates the evolution follows the $\{1,1,1\}$ Regge curve between $\alpha_1^{(011)}$ and $\alpha_2^{(111)}$. On the other hand, if $t_2^{(111)}$ is smaller than $t_1^{(011)}$, the evolution quickly crosses the $\{1,1,1\}$ Regge curve. In this case, the maximum mass of the subdominant mode can be estimated simply by $M_\text{s}^{(111)}(t_1^{(011)})$.

The above analysis applies to stages with other values of $l$. After identifying the time when the subdominant mode mass reaches its maximum, this maximum mass can be calculated using
%%%%%=====
\begin{align}
	\M[(1ll)]{M}{s,max} & \approx \M[(1ll)]{M}{s,0}\left[\frac{\M[(0ll)]{M}{s}(t_\text{2}^{(1ll)})}{\M[(0ll)]{M}{s,0}}\right]^{\beta_{1l}/\beta_{0l}},
\end{align}
%%%%%
with
%%%%%=====
\begin{align}
	\beta_{nl}\equiv\frac{(n+2l+1)!}{n!(n+l+1)^{2l+4}},
\end{align}
%%%%%
where $t_\text{max}^{(1ll)}$ is the greater one of $t_2^{(1ll)}$ and $t_1^{(0ll)}$.

After reaching the maximum mass, the subdominant mode enters the decaying phase and starts to shrink in size. We define $t_3^{(1ll)}$ as the time when the $\{1,l,l\}$ mode mass decreases to $1/e$ of its peak value. Two scenarios exist for estimating $t_3^{(1ll)}$. If the evolution does not follow the $\{0,l,l\}$ Regge curve, the accretion can be ignored. Then $t_3^{(1ll)}-t_1^{(0ll)}$ could be calculated in the same way as $\tau_\text{life}^{(111)}$ explained in Sec.~\ref{sec:evo_wo_acc}. The results with $l\leq 7$ are listed in Eqs.~\eqref{eq:life_1ll}. On the other hand, if the evolution follows the $\{0,l,l\}$ Regge curve, the growth of the BH mass is important. One starts from Eq.~\eqref{eq:Ms_ODE} with GW emission ignored. Using $a_*=a_{*\text{c}}^{(0ll)}$ given in Eq.~\eqref{eq:a_critc} and the exponential form of $M$ given in Eq.~\eqref{eq:BH_mass_efold}, one could then integrate Eq.~\eqref{eq:Ms_ODE} from $t_1^{(0ll)}$ to $t_3^{(1ll)}$ and obtain an expression similar to \eref{eq:Ms0p}. Then it is straightforward to calculate the mass coupling $\alpha_3^{(111)}$ at time $t_3^{(111)}$. Finally, one could solve for $t_3^{(111)}$ from the evolution of $M$ on the $\{0,l,l\}$ Regge curve. The calculation of the $M(t)$ on any $\{0,l,l\}$ Regge curve is explained in the previous subsection. For $l=1$, one arrives at Eq.~\eqref{eq:t_2} with $\alpha_2$ replaced by $\alpha_3^{(011)}$.

\subsection{GW emission}
\label{sec:Multi_GW}

We finally discuss the GW emission of each $l$-stage and the plausibility of observation by future GW interferometers. In the example presented in Fig.~\ref{fig:Multi}, one could observe that the GW emission flux in the $l=2$ and $l=3$ stages are comparable to that in the $l=1$ stage with accretion. It is very different from the case without accretion, in which the GW flux of the $l>1$ stage is many decades smaller \cite{Guo:2022mpr}.

\begin{figure*}[htbp]
	\centering 
	\includegraphics[width=0.9\textwidth]{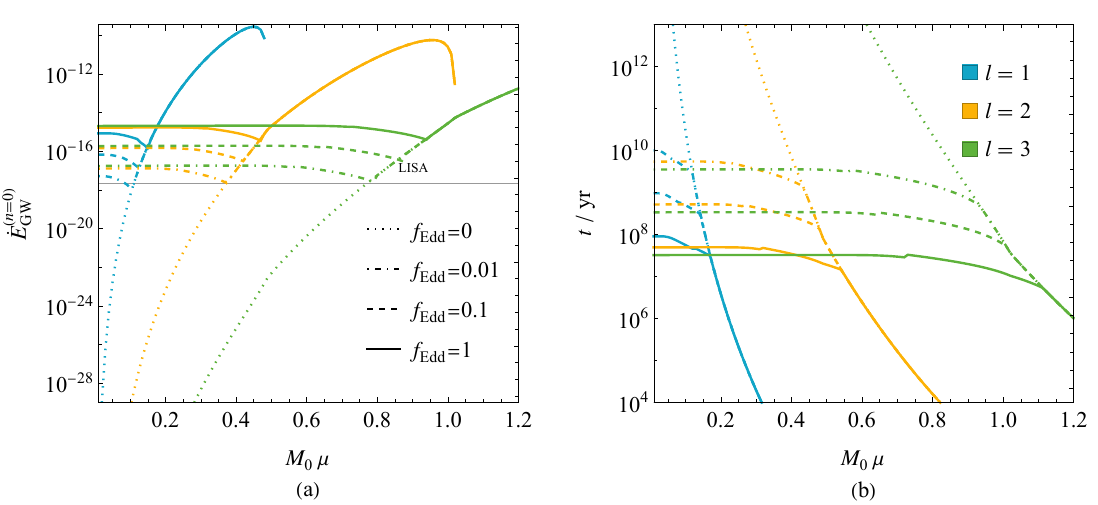}
	\includegraphics[width=0.9\textwidth]{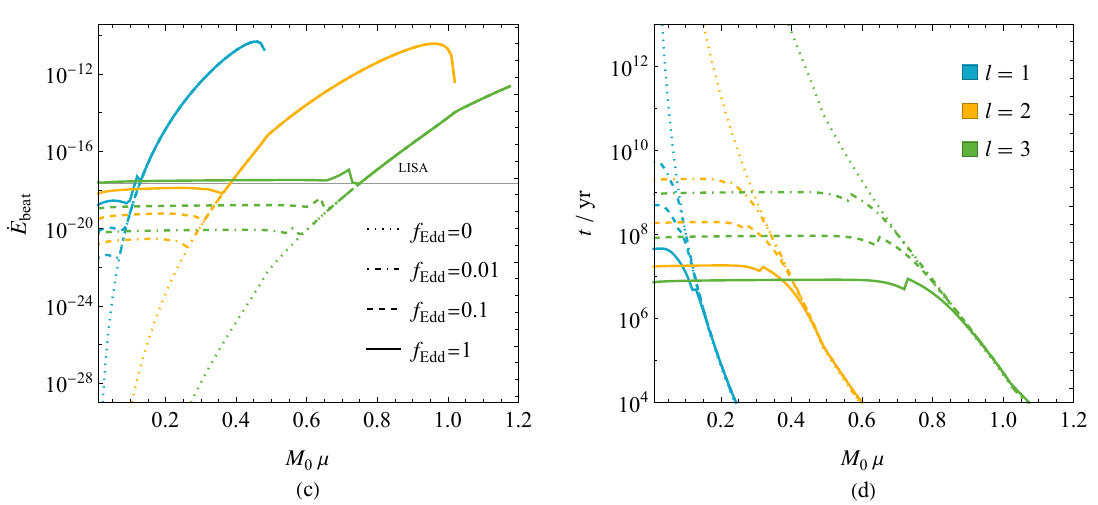}
	\caption{Panel (a) and (b): the GW fluxes $\dE[(0ll)]{E}{GW}$ and the time durations of the dominant modes as functions of the initial mass coupling. Panel (c) and (d): the GW beat signatures and their time durations as functions of the initial mass coupling. Curves with different colors are for different $l=m$ modes, indicated by the legends. The solid, dashed, dot-dashed, and dotted curves represent $\M[]{f}{Edd}=0,0.01,0.1$ and 1, respectively. Other parameters are the scalar mass $\mu = 10^{-17}$~eV, the initial BH spin $a_*=0.998$, and the initial condensate masses $\M[(nlm)]{M}{s,0}=10^{-5}M_0$. The horizontal lines in panels (a) and (c) illustrate the flux at which the characteristic strain equals the noise characteristic strain of the LISA, with the source located at redshift $z=0.1$ and the observation time as 4 years.} 
	\label{fig:vs_LISA} 
\end{figure*}

To study how accretion affects the GW emission with other parameters, the GW emission fluxes of the dominant modes with $l\leq 3$ and their timescales are plotted as functions of $M_0\mu$ in \fref{fig:vs_LISA} (a) and (b), with four different accretion rates $\M[]{f}{Edd}$. The analytical approximations of the BH mass $M$, cloud mass $M_\text{s}^{(0ll)}(t)$, and different time scales are utilized. We set the scalar mass as $\mu = 10^{-17}$ eV, the initial BH spin as $a_*=0.998$, and the initial condensate masses for all dominant and subdominant modes as $10^{-5}M_0$. In panel (a), the horizontal line denotes the flux at which the characteristic strain equals the noise characteristic strain of the LISA, with the source located at redshift $z = 0.1$, and the observation time as 4 years. The calculations are explained in \rscite{Brito:2017zvb,Guo:2022mpr}. 

From \fref{fig:vs_LISA} (a) and (b), one observes that accretion greatly enhances the GW emission flux, and at the same time reduces the GW duration time for small $M_0\mu$. With the BH mass $M$ fixed, the small-mass scalars which are undetectable via superradiance without accretion can be observed by LISA with $f_\text{Edd}$ as small as 0.01. This finding indicates future GW interferometers like LISA can search for a much larger range of the scalar mass than previously thought. Moreover, it is generally recognized in the literature that each BH undergoes boson annihilation only once in its cosmic history, as subsequent annihilation signals are significantly weaker \cite{Brito:2017zvb,Brito:2017wnc}. This is not generally true with accretion. On the one hand, a BH could have multiple boson annihilation in its cosmic history. For example, the lifetimes of the GW emission with $f_\text{Edd}=1$ and $\alpha_0 = 0.25$ are $2.18 \times 10^5\;\mathrm{yr}$, $4.94 \times 10^7\;\mathrm{yr}$ and $3.27 \times 10^7\;\mathrm{yr}$ for $l=1,2,3$, respectively. The numbers are $2.18 \times 10^5\;\mathrm{yr}$, $4.89 \times 10^8\;\mathrm{yr}$, and $3.41 \times 10^8\;\mathrm{yr}$ if $f_\text{Edd}$ is changed to 0.1. On the other hand, the GW emission flux with $l\geq 2$ modes could also be observed. This observation could be important in the study of stochastic GW background. We leave such phenomenological studies in future publications.

In \fref{fig:vs_LISA} (c) and (d), we further study the GW beat signal strength and its duration. The analytical approximations for the former are listed in App.~\ref{sec:Approximate_expressions}, with more details on GW beats given in \rcite{Guo:2022mpr}. The horizontal line in panel (c) is the same as that in panel (a). Same as the GW flux, the accretion also enhances the beat signal and reduces its duration in each $l$-stage. The beat signal without accretion has been carefully studied in Ref.~\cite{Guo:2022mpr}. With accretion, the enhanced beat signal could greatly facilitate the search for light scalars.

%%%%%FFFFF
%%%%%%

\section{Summary and Discussion}\label{sec:summary}

In this work, we provided a thorough analysis of how accretion affects the evolution of a BH-condensate system with scalar superradiance and GW emission. Despite the coexistence of multiple modes with different values of $n$ and $l$, the evolution has a simple and universal pattern. Our main findings are summarized below. In theoretical aspect: 
\begin{itemize}
	\item 
	We considered, for the first time, the capture of photons from the accretion disk by the BH during superradiant evolution. Despite the presence of superradiance, the BH spin still does not exceed its limit value $a_{*\text{lim}} \simeq 0.998$.
	\item With the small mass-coupling $\alpha = M\mu$, we obtained analytical approximations of all the important quantities, incorporating the effects of superradiance, accretion and GW emission. These formulas can be readily applied in relevant phenomenological studies.
	%Compared with numerical results, these approximations agree reasonably well (see Tables \ref{tab:single} and \ref{tab:Multi}).
	\item This work serves as a theoretical baseline for more complex scenarios, such as including non-gravitational interactions between the scalar field and Standard Model particles.
\end{itemize}
In phenomenological aspect, we found:
\begin{itemize}
	\item Accretion could greatly speed up the mass accumulation and the GW emission, resulting in stronger and shorter GW signal compared to the case without accretion. 
	\item GWs emitted by modes with $l > 1$ should also be taken into account when studying the stochastic GW background and conducting targeted GW searches, even though they are often neglected in the literature.
\end{itemize}

In Sec.~\ref{sec:setup}, we reviewed the scalar superradiance rate, the evolution of the BH-condensate system without accretion, and the evolution of BHs with only accretion. The capture of photons from the accretion disk is included, resulting in a limiting value of the BH spin $a_{*\text{lim}}=0.998$. Without superradiance, the BH mass grows exponentially with time and the spin follows the accretion trajectory to $a_{*\text{lim}}$. The differential evolution equations with all these effects are provided in Eqs.~\eqref{eq:total_ODEs}, which serves as the analysis benchmark in the rest of this manuscript.

In Sec.~\ref{sec:single}, we kept only the $\{0,1,1\}$ mode and study the influence of accretion on this BH-condensate system. Without accretion, the system has two characteristic time scales, i.e., the superradiance time scale $\tau_\text{SR}$ and the GW emission time scale $\tau_\text{GW}$. In the presence of accretion, there is an additional accretion time scale $\tau_\text{acc}$, given in Eq.~\eqref{eq:tau_acc}. Generally, the evolution could be split into the spin-up, spin-down, attractor, and decaying phases. In each phase, the unimportant effects could be ignored and analytical estimates of the important quantities have been obtained as a series of $\alpha$. When $\tau_\text{acc}\geq \tau_\text{SR}$, the evolution is strongly affected by the accretion, especially in the parameter space with small initial mass coupling or small initial BH spin. In particular, the condensate accumulates mass as a double exponential function, leading to a much stronger GW signal than that without accretion. The lifetime of the mode and the duration of the GW emission are greatly reduced, which is in agreement with the findings of Refs.~\cite{Brito:2014wla,Sarmah:2024nst}.

Then in Sec.~\ref{sec:multi}, we generalized the analysis of the single-mode case to that with multiple modes. The $\{n,l,l\}$ modes with $n=0,1$ and $0\leq l \leq 7$ are included. The evolution on the Regge plot presents a quite simple and universal pattern. It consists of the parts following along the accretion trajectory or different $\{0,l,l\}$ Regge trajectories, as well as the transitions between these curves, see Fig.~\ref{fig:Regge_Multi}. Analytical approximations of the start and end of each transition are provided, which agree reasonably well with the numerical results. Same as the single-mode case, the accretion impact strongly on those modes with superradiance time scale $\tau_\text{SR}$ equal to or larger than the accretion time scale $\tau_\text{acc}$. The accretion enhances the magnitude and reduces the duration of the GW emission from these modes.

In this work, we did not consider the back-reaction of the scalar condensate on the spacetime metric as well as the self-interaction of the scalar field. The self-interaction could generate much richer physics in the evolution, such as the bosonova collapse in Refs.~\cite{Yoshino:2012kn, Fukuda:2019ewf, Omiya:2022mwv,Omiya:2022gwu,Unal:2023yxt,Omiya:2024xlz}. Moreover, the coexistence of an accretion disk and a scalar condensate may lead to new observational signatures if their interactions exist. It has been shown that interaction between scalar and photons could modify the cosmic microwave background \cite{Blas:2020nbs} or result in a birefringent effect of the background light \cite{Chen:2019fsq,Chen:2021lvo}. 

Finally, we have assumed the dominant mode is always more massive than the subdominant one with the same azimuthal number $l$. Not considering the unnatural cases in which the subdominant mode has a much larger initial mass, this statement is strictly valid only for $l\leq 2$. For $l\geq 3$, there exists the ``overtone mixing" phenomenon which says the superradiance rate of the subdominant mode is larger than the dominant mode in some parameter space. Consequently, the subdominant mode is more massive even when its initial mass is not very large. Next, we argue the above approximations for the BH evolution could still give reasonable results. With overtone mixing, both the dominant and subdominant modes should be kept. Since $\omega_{0ll}\approx \omega_{1ll}$, the second piece on the right side of Eq.~\eqref{eq:J_BH_ODE} is roughly $-l \,\omega_{0ll}^{-1}\, (\dot{E}_\text{SR}^{(0ll)}+\dot{E}_\text{SR}^{(1ll)})$. The Eq.~\eqref{eq:Ms_ODE} for these two modes could also be added. The obtained differential equations are almost the same as Eqs.~\eqref{eq:total_ODEs}, except the $M_\text{s}^{(0ll)}$ and $\dot{E}_\text{SR}^{(0ll)}$ are replaced by $M_\text{s}^{(0ll)}+M_\text{s}^{(1ll)}$ and $\dot{E}_\text{SR}^{(0ll)}+\dot{E}_\text{SR}^{(1ll)}$, respectively. With these two replacements, the later analysis is unchanged, which results in the same evolution of the BH on the Regge plot. For the $l=4$ stage, the overtone mixing introduces an additional percentage error of roughly $10\%$ compared to the numerical calculation. Nonetheless, since the dominant and subdominant modes have very different GW emission rates, the GW emission flux will be largely modified with overtone mixing. We leave the related analysis in a future publication.

\section*{Acknowledgements}
T. Li is supported in part by the National Key Research and Development Program of China Grant No. 2020YFC2201504, by the Projects No. 11875062, No. 11947302, No. 12047503, and No. 12275333 supported by the National Natural Science Foundation of China, by the Key Research Program of the Chinese Academy of Sciences, Grant No. XDPB15, by the Scientific Instrument Developing Project of the Chinese Academy of Sciences, Grant No. YJKYYQ20190049, and by the International Partnership Program of Chinese Academy of Sciences for Grand Challenges, Grant No. 112311KYSB20210012. S-S. Bao and Y-D. Guo and H. Zhang are supported by the National Natural Science Foundation of China (Grants Nos. 12075136 and 124B2098) and the Natural Science Foundation of Shandong Province (Grant No. ZR2020MA094). 
\\

\appendix

\section{Some useful formulae}\label{sec:Approximate_expressions}

\noindent
{\bf (1) The energy and angular momentum of particles on the innermost stable circular orbit}
\\

For particles in the innermost stable circular orbit, the energy per mass $E_\mathrm{ms}^{\dagger}$ and angular momentum per mass $J_\mathrm{ms}^{\dagger}$ are given by \cite{Bardeen:1970zz,Bardeen:1972fi}
%%%%%=====
\begin{subequations}
	\begin{align}
		E_\mathrm{ms}^{\dagger} & = \sqrt{1 - \frac{2}{3r_\mathrm{ms}/M}},\\
		J_\mathrm{ms}^{\dagger} & = \frac{2M}{3\sqrt{3}}\left(1+2\sqrt{3r_\mathrm{ms}/M - 2}\right),
	\end{align}
\end{subequations}
%%%%%
where we take direct orbits, corotating with $J_\mathrm{ms}^{\dagger}>0$, and $r_\mathrm{ms}$ denotes the radius of the last stable circular orbit
%%%%%=====
\begin{align}
	r_\mathrm{ms}/M & = 3 + z_2 -\sqrt{(3-z_1) (3+z_1+2 z_2)},\\
	z_1 & = 1+\sqrt[3]{1-a_*^2} \left(\sqrt[3]{1+a_*} + \sqrt[3]{1-a_*}\right),\\
	z_2 & = \sqrt{3 a_*^2 + z_1^2}.
\end{align}
%%%%% 

\noindent
{\bf (2) Leading-order results of GW emission fluxes from dominant modes}
\\

The general formula can be written as
%%%%%=====
\begin{align}\label{eq:dot_E_GW}
	\dE[(0ll)]{E}{GW} = \M[(0ll)]{C}{GW} \left(\frac{\M[(0ll)]{M}{s}}{M}\right)^2 \left(\frac{M\mu}{l}\right)^{4l+10}.
\end{align}
%%%%%
For $l=1$ to 7, the corresponding coefficients are:
%%%%%=====
\begin{subequations}\label{eq:C_GW}
	\begin{align}
		\M[(011)]{C}{GW} & \approx 2.49\times 10^{-2}, \\
		\M[(022)]{C}{GW} & \approx 7.29\times10^{-2}, \\
		\M[(033)]{C}{GW} & \approx 8.97\times 10^{-2}, \\
		\M[(044)]{C}{GW} & \approx 4.29, \\
		\M[(055)]{C}{GW} & \approx 2.24\times10^2, \\
		\M[(066)]{C}{GW} & \approx 6.45\times 10^3, \\
		\M[(077)]{C}{GW} & \approx 1.46\times 10^5.
	\end{align}
\end{subequations}
%%%%%
%%%
\vskip 0.5cm
\noindent
{\bf (3) Leading-order results of the GW beat amplitude}
\\

The general formula is
%%%%%=====
\begin{align}
	\dE[(l)]{E}{beat} = \M[(l)]{C}{beat} \frac{\left(\M[(0ll)]{M}{s}\right)^{3/2}  \left(\M[(1ll)]{M}{s}\right)^{1/2}}{M^2} \left(\frac{M\mu}{l}\right)^{4l+10},
\end{align}
%%%%%
which is valid only when $M_\mathrm{s}^{(0ll)} > M^{(1ll)}_\mathrm{s}$. 
For $l=1$ to 7, the corresponding coefficients are
%%%%%=====
\begin{subequations}
	\begin{align}
		\M[(1)]{C}{beat} & \approx 5.89\times 10^{-2}, \\
		\M[(2)]{C}{beat} & \approx 2.26\times 10^{-1}, \\
		\M[(3)]{C}{beat} & \approx 3.33\times 10^{-1}, \\
		\M[(4)]{C}{beat} & \approx 1.82\times 10^1,\\
		\M[(5)]{C}{beat} & \approx 1.06\times 10^3,\\
		\M[(6)]{C}{beat} & \approx 3.11\times 10^4,\\
		\M[(7)]{C}{beat} & \approx 8.11\times 10^5.
	\end{align}
\end{subequations}
%%%%%
%%%
\vskip 0.5cm
\noindent
{\bf (4) Leading-order results of subdominant mode lifetimes}
\\

The general formula is
%%%%%=====
\begin{align}\label{eq:life_1ll}
	\frac{\M[(1ll)]{\tau}{life}}{M} = \M[(1ll)]{C}{life}  \left(\frac{M\mu}{l}\right)^{-4l-8}.
\end{align}
%%%%%
For $l=1$ to 7, the corresponding coefficients are
%%%%%=====
\begin{subequations}
	\begin{align}
		\M[(111)]{C}{life} & \approx 1.63\times 10^2, \\
		\M[(122)]{C}{life} & \approx 3.88, \\
		\M[(133)]{C}{life} & \approx 4.29\times	10^{-1}, \\
		\M[(144)]{C}{life} & \approx 8.26\times 10^{-2}, \\
		\M[(155)]{C}{life} & \approx 2.10\times 10^{-2}, \\
		\M[(166)]{C}{life} & \approx 6.27\times 10^{-3}, \\
		\M[(177)]{C}{life} & \approx 2.09\times 10^{-3}.
	\end{align}
\end{subequations}
%%%%%
%%%
\vskip 0.5cm
\noindent
{\bf (5) The condensate mass when the system follows the accretion trajectory}
\\

The general formula is given by:
%%%%%=====
\begin{align}\label{eq:Ms_accretion_trajectory}
	M^{(0ll)}_\text{s} = M^{(0ll)}_\text{s0} \exp\left\{\alpha_0^{(4l+4)}\mu\tau_\text{acc}\left[g_{0ll}\left(\frac{M}{M_0}\right) - g_{0ll}\left(1\right)\right]\right\}, \hspace{0.5cm} \text{for }  1 \leq \frac{M}{M_0} \leq {\sqrt{6}\,k}.&
\end{align}
%%%%%
For $l=1$ to 7, the corresponding expressions for $g_{0ll}$ are
%%%%%=====
\begin{subequations}
	\begin{align}
		g_{011}(x) \equiv & \frac{1}{36} \left[\frac{1}{35} \sqrt{3} k \left(216 k^4+36 k^2 x^2+5 x^4\right) z^{3}+\frac{2}{7} \sqrt{6} k x^7-\frac{2 \alpha _0 x^9}{3}\right],\\
		\begin{split}
			g_{022}(x) \equiv & -1.11\times 10^{-5} \alpha_0 x^{13} + k \left(2.15\times 10^{-5} - 3.68\times 10^{-4} k \alpha_0\right) x^{11} 
			\\
			&+ k^2 \left(2.97\times 10^{-4} z \alpha_0 - 7.58\times 10^{-6}\right) x^{10} + k^3 \left(5.24\times 10^{-4} - 5.78\times 10^{-4} k \alpha_0\right) x^9
			\\
			&+ k^4 \left(-2.97\times 10^{-4} z \alpha_0 - 8.45\times 10^{-4}\right) x^8 + 3.64\times 10^{-3} k^5 x^7 
			\\
			&+ k^6 \left(7.45\times 10^{-4} - 3.06\times 10^{-3} z \alpha_0\right) x^6 + k^7 z \left(1.35\times 10^{-2} - 3.3\times 10^{-2} k \alpha_0\right) x^4 
			\\
			&+ k^9 z \left(1.61\times 10^{-1} - 3.96\times 10^{-1} k \alpha_0\right) x^2 + k^{11} z \left(2.91 - 7.13 k \alpha_0\right)
			,
		\end{split}
		\\
		\begin{split}
			g_{033}(x) \equiv & -1.82\times 10^{-9} \alpha_0 x^{17} + k \left(5.06\times 10^{-9} - 1.78\times 10^{-7} k \alpha_0\right) x^{15} 
			\\
			&+ k^2 \left(1.44\times 10^{-7} z \alpha_0 - 1.79\times 10^{-9}\right) x^{14} + k^3 \left(3.6\times 10^{-7} - 9.85\times 10^{-7} k \alpha_0\right) x^{13} 
			\\
			&+ k^4 \left(3.26\times 10^{-6} z \alpha_0 - 5.85\times 10^{-7}\right) x^{12} + k^5 \left(9.49\times 10^{-7} - 2.07\times 10^{-5} k \alpha_0\right) x^{11} 
			\\
			&+ k^6 \left(1.37\times 10^{-6} z \alpha_0 - 1.02\times 10^{-5}\right) x^{10} + k^7 \left(9.45\times 10^{-5} - 4.96\times 10^{-6} k \alpha_0\right) x^9
			\\
			& + k^8 \left(-4.24\times 10^{-5} z \alpha_0 - 2.16\times 10^{-5}\right) x^8 + 7.81\times 10^{-5} k^9 x^7 
			\\
			&+ k^{10} \left(2.03\times 10^{-4} - 4.36\times 10^{-4} z \alpha_0\right) x^6 + k^{11} z \left(2.26\times 10^{-3} - 4.71\times 10^{-3} k \alpha_0\right) x^4  
			\\
			&+ k^{13} z \left(2.72\times 10^{-2} - 5.65\times 10^{-2} k \alpha_0\right) x^2 + k^{15} z \left(4.89\times 10^{-1} - 1.02 k \alpha_0\right)
			,
		\end{split}
		\\
		\begin{split}
			g_{044}(x) \equiv & -1.15\times 10^{-13} \alpha _0 x^{21}+k \left(4.15\times 10^{-13}-2.23\times 10^{-11} k \alpha _0\right) x^{19}
			\\
			&+k^2 \left(1.8\times 10^{-11} z \alpha _0-1.47\times 10^{-13}\right) x^{18} +k^3 \left(5.81\times 10^{-11}-2.85\times 10^{-10} k \alpha _0\right) x^{17}
			\\
			&+k^4 \left(1.17\times 10^{-9} z \alpha _0-9.44\times 10^{-11}\right) x^{16} + k^5 \left(-2.6\times 10^{-9} k \alpha _0-2.51\times 10^{-10}\right) x^{15}
			\\
			&+k^6 \left(9.61\times 10^{-9} z \alpha _0-4.68\times 10^{-9}\right) x^{14} +k^7 \left(1.2\times 10^{-8}-1.02\times 10^{-7} k \alpha _0\right) x^{13}
			\\
			&+k^8 \left(4.31\times 10^{-8} z \alpha _0-3.89\times 10^{-8}\right) x^{12} +k^9 \left(4.62\times 10^{-7}-1.73\times 10^{-7} k \alpha _0\right) x^{11}
			\\
			&+k^{10} \left(-1.69\times 10^{-7} z \alpha _0-2.91\times 10^{-7}\right) x^{10} +k^{11} \left(1.53\times 10^{-6}-1.62\times 10^{-8} k \alpha _0\right) x^9
			\\
			&+k^{12} \left(7.46\times 10^{-7}-1.97\times 10^{-6} z \alpha _0\right) x^8 +4.77\times 10^{-7} k^{13} x^7
			\\
			&+k^{14} \left(1.16\times 10^{-5}-2.03\times 10^{-5} z \alpha _0\right) x^6+k^{15} z \left(1.26\times 10^{-4}-2.19\times 10^{-4} k \alpha _0\right) x^4
			\\
			&+k^{17} z \left(1.51\times 10^{-3}-2.62\times 10^{-3} k \alpha _0\right) x^2+k^{19} z \left(2.72\times 10^{-2}-4.72\times 10^{-2} k \alpha _0\right)
			,
		\end{split}
		\\
		\begin{split}
			g_{055}(x) \equiv & -3.38\times 10^{-18} \alpha _0 x^{25}+k \left(1.5\times 10^{-17}-1.08\times 10^{-15} k \alpha _0\right) x^{23}
			\\
			&+k z \left(8.75\times 10^{-16} k \alpha _0-5.3\times 10^{-18}\right) x^{22} + k^3 \left(3.46\times 10^{-15}-2.53\times 10^{-14} k \alpha _0\right) x^{21}
			\\
			&+k^4 \left(1.1\times 10^{-13} z \alpha _0-5.62\times 10^{-15}\right) x^{20} + k^5 \left(4.84\times 10^{-13} k \alpha _0-4.87\times 10^{-14}\right) x^{19}
			\\
			&+k^6 \left(2.32\times 10^{-12} z \alpha _0-5.37\times 10^{-13}\right) x^{18} + k^7 \left(-1.31\times 10^{-11} k \alpha _0-4.27\times 10^{-12}\right) x^{17}
			\\
			&+k^8 \left(1.79\times 10^{-11} z \alpha _0-9.76\times 10^{-12}\right) x^{16} +k^9 \left(6.77\times 10^{-11}-1.63\times 10^{-10} k \alpha _0\right) x^{15}
			\\
			&+k^9 z \left(1.32\times 10^{-10} k \alpha _0-1.04\times 10^{-10}\right) x^{14} +k^{11} \left(8.35\times 10^{-10}-8.03\times 10^{-10} k \alpha _0\right) x^{13}
			\\
			&+k^{12} \left(-1.41\times 10^{-10} z \alpha _0-8.63\times 10^{-10}\right) x^{12} +k^{13} \left(6.29\times 10^{-9}-4.9\times 10^{-10} k \alpha _0\right) x^{11}
			\\
			&+k^{14} \left(-4.01\times 10^{-9} z \alpha _0-1.15\times 10^{-10}\right) x^{10} +k^{15} \left(7.07\times 10^{-9}-2.45\times 10^{-11} k \alpha _0\right) x^9
			\\
			&+k^{16} \left(2.61\times 10^{-8}-4.07\times 10^{-8} z \alpha _0\right) x^8+1.16\times 10^{-9} k^{17} x^7
			\\
			&+k^{18} \left(2.82\times 10^{-7}-4.18\times 10^{-7} z \alpha _0\right) x^6+k^{19} z \left(3.04\times 10^{-6}-4.52\times 10^{-6} k \alpha _0\right) x^4
			\\
			&+k^{21} z \left(3.65\times 10^{-5}-5.42\times 10^{-5} k \alpha _0\right) x^2+k^{23} z \left(6.57\times 10^{-4}-9.76\times 10^{-4} k \alpha _0\right)
			,
		\end{split}
		\\
		\begin{split}
			g_{066}(x) \equiv & -5.24\times 10^{-23} \alpha _0 x^{29}+k \left(2.76\times 10^{-22}-2.5\times 10^{-20} k \alpha _0\right) x^{27}
			\\
			&+k z \left(2.02\times 10^{-20} k \alpha _0-9.75\times 10^{-23}\right) x^{26} +k^3 \left(9.47\times 10^{-20}-9.3\times 10^{-19} k \alpha _0\right) x^{25}
			\\
			&+k^4 \left(4.16\times 10^{-18} z \alpha _0-1.54\times 10^{-19}\right) x^{24}+k^5 \left(5.65\times 10^{-17} k \alpha _0-2.53\times 10^{-18}\right) x^{23}
			\\
			&+k^6 \left(1.63\times 10^{-16} z \alpha _0-2.39\times 10^{-17}\right) x^{22}+k^7 \left(5.08\times 10^{-16} k \alpha _0-5.38\times 10^{-16}\right) x^{21}
			\\
			&+k^8 \left(1.93\times 10^{-15} z \alpha _0-7.64\times 10^{-16}\right) x^{20}+k^9 \left(-1.65\times 10^{-14} k \alpha _0-3.51\times 10^{-15}\right) x^{19}
			\\
			&+k^{10} \left(2.5\times 10^{-14} z \alpha _0-9.49\times 10^{-15}\right) x^{18}+k^{11} \left(7.75\times 10^{-14}-1.57\times 10^{-13} k \alpha _0\right) x^{17}
			\\
			&+k^{12} \left(1.56\times 10^{-13} z \alpha _0-1.69\times 10^{-13}\right) x^{16} +k^{13} \left(1.03\times 10^{-12}-1.19\times 10^{-12} k \alpha _0\right) x^{15}
			\\
			&+k^{14} \left(1.99\times 10^{-13} z \alpha _0-1.14\times 10^{-12}\right) x^{14} +k^{15} \left(9.33\times 10^{-12}-2.\times 10^{-12} k \alpha _0\right) x^{13}
			\\
			&+k^{16} \left(-3.97\times 10^{-12} z \alpha _0-2.95\times 10^{-12}\right) x^{12} + k^{17} \left(2.36\times 10^{-11}-6.26\times 10^{-13} k \alpha _0\right) x^{11}
			\\
			&+k^{18} \left(2.64\times 10^{-11}-4.35\times 10^{-11} z \alpha _0\right) x^{10} +k^{19} \left(1.34\times 10^{-11}-1.94\times 10^{-14} k \alpha _0\right) x^9
			\\
			&+k^{20} \left(3.35\times 10^{-10}-4.36\times 10^{-10} z \alpha _0\right) x^8+1.34\times 10^{-12} k^{21} x^7
			\\
			&+k^{22} \left(3.47\times 10^{-9}-4.48\times 10^{-9} z \alpha _0\right) x^6+k^{23} z \left(3.74\times 10^{-8}-4.84\times 10^{-8} k \alpha _0\right) x^4
			\\
			&+k^{25} z \left(4.49\times 10^{-7}-5.81\times 10^{-7} k \alpha _0\right) x^2+k^{27} z \left(8.08\times 10^{-6}-1.05\times 10^{-5} k \alpha _0\right)
			,
		\end{split}
		\\
		\begin{split}
			g_{077}(x) \equiv & -4.69\times 10^{-28} \alpha_0 x^{33} + k \left( 2.85\times 10^{-27} - 3.13\times 10^{-25} k \alpha_0 \right) x^{31} 
			\\
			&+ k^2 \left( 2.53\times 10^{-25} z \alpha_0 - 1.01\times 10^{-27} \right) x^{30} + k^3 \left( 1.36\times 10^{-24} - 1.69\times 10^{-23} k \alpha_0 \right) x^{29} 
			\\
			&+ k^4 \left( 7.67\times 10^{-23} z \alpha_0 - 2.22\times 10^{-24} \right) x^{28} + k^5 \left( 1.94\times 10^{-21} k \alpha_0 - 5.76\times 10^{-23} \right) x^{27} 
			\\
			&+ k^6 \left( 4.78\times 10^{-21} z \alpha_0 - 5.07\times 10^{-22} \right) x^{26} + k^7 \left( 7.37\times 10^{-20} k \alpha_0 - 2.09\times 10^{-20} \right) x^{25}  
			\\
			&+ k^8 \left( 8.34\times 10^{-20} z \alpha_0 - 2.52\times 10^{-20} \right) x^{24} + k^9 \left( -2.36\times 10^{-20} k \alpha_0 - 5.43\times 10^{-19} \right) x^{23}  
			\\
			&+ k^{10} \left( 1.18\times 10^{-18} z \alpha_0 - 3.57\times 10^{-19} \right) x^{22} + k^{11} \left( -7.24\times 10^{-18} k \alpha_0 - 7.36\times 10^{-19} \right) x^{21}  
			\\
			&+ k^{12} \left( 1.85\times 10^{-17} z \alpha_0 - 7.52\times 10^{-18} \right) x^{20} + k^{13} \left( 2.94\times 10^{-17} - 1.15\times 10^{-16} k \alpha_0 \right) x^{19}  
			\\
			&+ k^{14} \left( 1.12\times 10^{-16} z \alpha_0 - 1.33\times 10^{-16} \right) x^{18} + k^{15} \left( 8.8\times 10^{-16} - 8.31\times 10^{-16} k \alpha_0 \right) x^{17}
			\\
			& + k^{15} z \left( 3.39\times 10^{-16} k \alpha_0 - 9.62\times 10^{-16} \right) x^{16} + k^{17} \left( 7.05\times 10^{-15} - 2.63\times 10^{-15} k \alpha_0 \right) x^{15}
			\\
			&+ k^{18} \left( -1.84\times 10^{-15} z \alpha_0 - 3.78\times 10^{-15} \right) x^{14}  =+ k^{19} \left( 2.92\times 10^{-14} - 2.19\times 10^{-15} k \alpha_0 \right) x^{13} 
			\\
			&+ k^{20} \left( 1.06\times 10^{-14} - 2.69\times 10^{-14} z \alpha_0 \right) x^{12} + k^{21} \left( 3.64\times 10^{-14} - 4.18\times 10^{-16} k \alpha_0 \right) x^{11} 
			\\
			& + k^{21} z \left( 2.25\times 10^{-13} - 2.68\times 10^{-13} k \alpha_0 \right) x^{10} + k^{23} \left( 1.24\times 10^{-14} - 8.81\times 10^{-18} k \alpha_0 \right) x^9
			\\
			&+ k^{23} z \left( 2.34\times 10^{-12} - 2.68\times 10^{-12} k \alpha_0 \right) x^8 + 8.42\times 10^{-16} k^{25} x^7 
			\\
			&+ k^{26} \left( 2.4\times 10^{-11} - 2.75\times 10^{-11} z \alpha_0 \right) x^6 + k^{27} z \left( 2.6\times 10^{-10} - 2.97\times 10^{-10} k \alpha_0 \right) x^4 
			\\
			&+ k^{29} z \left( 3.11\times 10^{-9} - 3.57\times 10^{-9} k \alpha_0 \right) x^2 + k^{31} z \left( 5.61\times 10^{-8} - 6.42\times 10^{-8} k \alpha_0 \right)
			,
		\end{split}
	\end{align}
\end{subequations}
%%%%%
where $z\equiv\sqrt{9 k^2-x^2}$.
%%%
\vskip 0.5cm
\noindent

{\bf (6) Approximate formulae for the mass couplings when the system leaves the accretion trajectory and merges onto the $\{0,l,l\}$ Regge trajectory}
\\

If only accretion and the $\{0,l,l\}$ mode exist, the mass couplings when the system leaves the accretion trajectory can be estimated by:
%===
\begin{subequations}\label{eq:alpha_0p_0ll}
	\begin{align}
		\alpha_{0'}^{(011)} &= \left(\frac{2^4 3^2}{\tau_{\mathrm{acc}}\mu}\right)^{{1}/{8}}
		\left[
		W_0\left(2^{{4}/{9}}3^{{2}/{3}} M_\text{s,0}^{-{4}/{3}} \tau_\text{acc}^{-{1}/{3}}\mu^{-{5}/{3}}e^{\frac{\alpha_{0}^8\tau_{\mathrm{acc}}\mu}{2^4 3^2}}
		\right)
		\right]^{{1}/{8}},
		\\
		\alpha_{0'}^{(022)}&=\left(\frac{3^{11}5^{2}}{2^{7}\tau_\mathrm{acc}\mu}\right)^{1/12}\left[W_0\left(2^{-17/5}3^{17/5}5^{-4/5}M_\mathrm{s,0}^{-6/5}\tau_\mathrm{acc}^{-1/5}\mu^{-7/5}e^{\frac{2^{7}\alpha_0^{12}\tau_\mathrm{acc}\mu}{3^{11}5^{2}}}\right)\right]^{1/12},
		\\
		\alpha_{0'}^{(033)}&=\left(\frac{2^{17}5^{3}7^{2}}{3^{3}\tau_\mathrm{acc}\mu}\right)^{1/16}\left[W_0\left(2^{5/3}3^{-3/7}5^{3/7}7^{-6/7}M_\mathrm{s,0}^{-8/7}\tau_\mathrm{acc}^{-1/7}\mu^{-9/7}e^{\frac{3^{3}\alpha_0^{16}\tau_\mathrm{acc}\mu}{2^{17}5^{3}7^{2}}}\right)\right]^{1/16},
		\\
		\begin{split}
			\alpha_{0'}^{(044)}&=\left(\frac{3^{8}5^{15}7^{3}}{2^{20}\tau_\mathrm{acc}\mu}\right)^{1/20}
			\\
			& \hspace{0.5cm} \times \left[W_0\left(2^{-140/27}3^{-2/9}5^{5/3}7^{1/3}M_\mathrm{s,0}^{-10/9}\tau_\mathrm{acc}^{-1/9}\mu^{-11/9}e^{\frac{2^{20}\alpha_0^{20}\tau_\mathrm{acc}\mu}{3^{8}5^{15}7^{3}}}\right)\right]^{1/20},
		\end{split}
		\\
		\begin{split}
			\alpha_{0'}^{(055)}&=\left(\frac{2^{11}3^{24}7^{3}11^{2}}{5^{7}\tau_\mathrm{acc}\mu}\right)^{1/24}
			\\
			& \hspace{0.5cm} \times\left[W_0\left(2^{3/11}3^{36/11}5^{-19/11}7^{3/11}11^{-10/11}M_\mathrm{s,0}^{-12/11}\tau_\mathrm{acc}^{-1/11}\mu^{-13/11}e^{\frac{5^{7}\alpha_0^{24}\tau_\mathrm{acc}\mu}{2^{11}3^{24}7^{3}11^{2}}}\right)\right]^{1/24},
		\end{split}
		\\
		\begin{split}
			\alpha_{0'}^{(066)}&=\left(\frac{5^{4}7^{19}11^{3}13^{2}}{2^{16}3^{2}\tau_\mathrm{acc}\mu}\right)^{1/28}
			\\
			& \hspace{0.5cm} \times\left[W_0\left(\frac{5^{4/13}7^{19/13}11^{3/13}}{2^{118/39}3^{2/13}13^{12/13}}M_\mathrm{s,0}^{-14/13}\tau_\mathrm{acc}^{-1/13}\mu^{-15/13}e^{\frac{2^{16}3^{2}\alpha_0^{28}\tau_\mathrm{acc}\mu}{5^{4}7^{19}11^{3}13^{2}}}\right)\right]^{1/28},
		\end{split}
		\\
		\begin{split}
			\alpha_{0'}^{(077)}&=\left(\frac{2^{50}3^{13}5^{6}11^{3}13^{3}}{7^{11}\tau_\mathrm{acc}\mu}\right)^{1/32}
			\\
			& \hspace{0.5cm} \times\left[W_0\left(\frac{2^{118/45}3^{13/15}11^{1/5}13^{1/5}}{5^{2/3}7^{9/5}}M_\mathrm{s,0}^{-16/15}\tau_\mathrm{acc}^{-1/15}\mu^{-17/15}e^{\frac{7^{11}\alpha_0^{32}\tau_\mathrm{acc}\mu}{2^{50}3^{13}5^{6}11^{3}13^{3}}}\right)\right]^{1/32}.
		\end{split}
	\end{align}
\end{subequations}
%===
The corresponding mass couplings when merges onto the $\{0,l,l\}$ Regge trajectory are given by:
%===
\begin{subequations}\label{eq:alpha_1_0lclc}
\begin{align}
	\alpha_{1}^{(011)}&=\left[-\frac{2^{4}3}{\tau_\mathrm{acc}\mu}W_{-1}\left(-\frac{M_\mathrm{s0'}^{4}\tau_\mathrm{acc}\mu^{5}a_{*0'}^{-4}}{2^{4}3}e^{-\frac{\alpha_{0'}^{8}\tau_\mathrm{acc}\mu}{2^{4}3}}\right)\right]^{1/8},
	\\
	\alpha_{1}^{(022)}&=\left[-\frac{2^{-7}3^{11}5}{\tau_\mathrm{acc}\mu}W_{-1}\left(-\frac{M_\mathrm{s0'}^{6}\tau_\mathrm{acc}\mu^{7}a_{*0'}^{-6}}{2^{-13}3^{11}5}e^{-\frac{\alpha_{0'}^{12}\tau_\mathrm{acc}\mu}{2^{-7}3^{11}5}}\right)\right]^{1/12},
	\\
	\alpha_{1}^{(033)}&=\left[-\frac{2^{17}3^{-3}5^{3}7}{\tau_\mathrm{acc}\mu}W_{-1}\left(-\frac{M_\mathrm{s0'}^{8}\tau_\mathrm{acc}\mu^{9}a_{*0'}^{-8}}{2^{17}3^{-11}5^{3}7}e^{-\frac{\alpha_{0'}^{16}\tau_\mathrm{acc}\mu}{2^{17}3^{-3}5^{3}7}}\right)\right]^{1/16},
	\\
	\alpha_{1}^{(044)}&=\left[-\frac{2^{-20}3^{6}5^{15}7^{3}}{\tau_\mathrm{acc}\mu}W_{-1}\left(-\frac{M_\mathrm{s0'}^{10}\tau_\mathrm{acc}\mu^{11}a_{*0'}^{-10}}{2^{-40}3^{6}5^{15}7^{3}}e^{-\frac{\alpha_{0'}^{20}\tau_\mathrm{acc}\mu}{2^{-20}3^{6}5^{15}7^{3}}}\right)\right]^{1/20},
	\\
	\alpha_{1}^{(055)}&=\left[-\frac{2^{11}3^{24}5^{-7}7^{3}11}{\tau_\mathrm{acc}\mu}W_{-1}\left(-\frac{M_\mathrm{s0'}^{12}\tau_\mathrm{acc}\mu^{13}a_{*0'}^{-12}}{2^{11}3^{24}5^{-19}7^{3}11}e^{-\frac{\alpha_{0'}^{24}\tau_\mathrm{acc}\mu}{2^{11}3^{24}5^{-7}7^{3}11}}\right)\right]^{1/24},
	\\
	\alpha_{1}^{(066)}&=\left[-\frac{2^{-16}3^{-2}5^{4}7^{19}11^{3}13}{\tau_\mathrm{acc}\mu}W_{-1}\left(-\frac{M_\mathrm{s0'}^{14}\tau_\mathrm{acc}\mu^{15}a_{*0'}^{-14}}{2^{-30}3^{-16}5^{4}7^{19}11^{3}13}e^{-\frac{\alpha_{0'}^{28}\tau_\mathrm{acc}\mu}{2^{-16}3^{-2}5^{4}7^{19}11^{3}13}}\right)\right]^{1/28},
	\\
	\alpha_{1}^{(077)}&=\left[-\frac{2^{50}3^{12}5^{5}7^{-11}11^{3}13^{3}}{\tau_\mathrm{acc}\mu}W_{-1}\left(-\frac{M_\mathrm{s0'}^{16}\tau_\mathrm{acc}\mu^{17}a_{*0'}^{-16}}{2^{50}3^{12}5^{5}7^{-27}11^{3}13^{3}}e^{-\frac{\alpha_{0'}^{32}\tau_\mathrm{acc}\mu}{2^{50}3^{12}5^{5}7^{-11}11^{3}13^{3}}}\right)\right]^{1/32}.
\end{align}
\end{subequations}
%%%%%

\vskip 0.5cm
\noindent
{\bf (7) Approximate formulae for superradiant energy fluxes of dominant modes in attractor phases}
\\

The general formula for the first two orders is given by:
%%%%%=====
\begin{align}
	\dE[(0ll)]{E}{SR} = \frac{\alpha  M}{\M[]{\tau}{acc}} 
	\left[ 
		\frac{3}{l} \sqrt{\frac32}
		-\frac{31}{2l^2} \alpha
		+\mathcal{O}(\alpha^2) 
	\right].
\end{align}
%%%%%
For $l=1$ to 7, the approximate formulae including higher order  are provided:
%%%%%=====
\begin{subequations}
	\begin{align}
		\dE[(011)]{E}{SR} & = 
		\frac{\alpha  M}{\M[]{\tau}{acc}} \left[ 3\sqrt{\frac32}-\frac{31\alpha}2+\frac{787\alpha^2}{8\sqrt{6}}-\frac{20305\alpha^3}{216} +\frac{1564549\alpha^4}{3456\sqrt{6}}+\mathcal{O}(\alpha^5) \right], \\
		\dE[(022)]{E}{SR} & = 
		\frac{\alpha  M}{\M[]{\tau}{acc}} \left[ \frac{3\sqrt{\frac32}}2-\frac{31\alpha}8+\frac{65}{16}\sqrt{\frac32}\alpha^2-\frac{9743\alpha^3}{1728}+\frac{223999\alpha^4}{17280\sqrt{6}} +\mathcal{O}(\alpha^5) \right], \\
		\dE[(033)]{E}{SR} & = 
		\frac{\alpha  M}{\M[]{\tau}{acc}} \left[ \sqrt{\frac32}-\frac{31\alpha}{18}+\frac{3103\alpha^2}{864\sqrt6}-\frac{75955\alpha^3}{69984}+\frac{151571899\alpha^4}{94058496\sqrt6} +\mathcal{O}(\alpha^5) \right], \\
		\dE[(044)]{E}{SR} & = 
		\frac{\alpha  M}{\M[]{\tau}{acc}} \left[ \frac{3\sqrt{\frac32}}4-\frac{31\alpha}{32}+\frac{4831\alpha^2}{3200\sqrt{6}}-\frac{233279\alpha^3}{691200}+\frac{5079739\alpha^4}{13824000\sqrt{6}} +\mathcal{O}(\alpha^5) \right], \\
		\dE[(055)]{E}{SR} & =
		\frac{\alpha  M}{\M[]{\tau}{acc}} \left[ \frac{3\sqrt{\frac32}}5-\frac{31\alpha}{50}+\frac{257\sqrt{\frac32}\alpha^2}{1000}-\frac{18433\alpha^3}{135000}+\frac{13894871\alpha^4}{118800000\sqrt{6}} +\mathcal{O}(\alpha^5) \right], \\
		\dE[(066)]{E}{SR} & = 
		\frac{\alpha  M}{\M[]{\tau}{acc}} \left[ \frac{\sqrt{\frac32}}2-\frac{31\alpha}{72}+\frac{9427\alpha^2}{21168\sqrt{6}}-\frac{447455\alpha^3}{6858432}+\frac{1205216951\alpha^4}{26212927104\sqrt{6}} +\mathcal{O}(\alpha^5) \right], \\
		\dE[(077)]{E}{SR} & = 
		\frac{\alpha  M}{\M[]{\tau}{acc}} \left[ \frac{3\sqrt{\frac32}}7-\frac{31\alpha}{98}+\frac{12295\alpha^2}{43904\sqrt{6}}-\frac{290131\alpha^3}{8297856}+\frac{1554440081\alpha^4}{74348789760\sqrt{6}} +\mathcal{O}(\alpha^5) \right].
	\end{align}
\end{subequations}
%%%%%

%%%
\vskip 0.5cm
\noindent
{\bf (8) Approximate formulae for the condensate mass of dominant modes in attractor phases}
\\

The general formula for the first two orders is given by:
%%%%%=====
\begin{align}
	\M[(0ll)]{M}{s,Regge} = \frac{\alpha^2}{\alpha_0} 
	\left[ 
		\frac{3}{2l} \sqrt{\frac32}
		-\frac{2}{3l^2} \alpha
		+\mathcal{O}(\alpha^2) 
	\right].
\end{align}
%%%%%
For $l=1$ to 7, the approximate formulae including higher order are provided:
\begin{subequations}\label{eq:Ms_Regge_0ll}
	\begin{align}
		\M[(011)]{M}{s,Regge} &= \frac{\alpha^2}{\alpha_0}\left(\frac{3}{2}\sqrt{\frac{3}{2}}	-\frac{2 \alpha }{3}	-\frac{473 \alpha ^2}{32 \sqrt{6}}	-\frac{223 \alpha ^3}{270} + \frac{999421 \alpha ^4}{20736 \sqrt{6}}\right),
		\\
		\M[(022)]{M}{s,Regge} &= \frac{\alpha^2}{\alpha_0}\left(\frac{3}{4} \sqrt{\frac{3}{2}}	-\frac{\alpha }{6}	-\frac{5}{8} \sqrt{\frac{3}{2}} \alpha ^2	-\frac{\alpha ^3}{27} + \frac{1313 \alpha ^4}{810 \sqrt{6}}\right),
		\\
		\M[(033)]{M}{s,Regge} &= \frac{\alpha^2}{\alpha_0}\left(\frac{1}{2}\sqrt{\frac{3}{2}}	-\frac{2 \alpha }{27}	-\frac{1937 \alpha ^2}{3456 \sqrt{6}}	-\frac{487 \alpha ^3}{87480} + \frac{125563483 \alpha ^4}{564350976 \sqrt{6}}\right),
		\\
		\M[(044)]{M}{s,Regge} &= \frac{\alpha^2}{\alpha_0}\left(\frac{3}{8}\sqrt{\frac{3}{2}}	-\frac{\alpha }{24}	-\frac{761 \alpha ^2}{3200 \sqrt{6}}	-\frac{151 \alpha ^3}{108000} + \frac{561893 \alpha ^4}{10368000 \sqrt{6}}\right),
		\\
		\M[(055)]{M}{s,Regge} &= \frac{\alpha^2}{\alpha_0}\left(\frac{3}{10}\sqrt{\frac{3}{2}}	-\frac{2 \alpha }{75}	-\frac{163 \sqrt{\frac{3}{2}} \alpha ^2}{4000}	-\frac{79 \alpha ^3}{168750}	+ \frac{2577331 \alpha ^4}{142560000 \sqrt{6}}\right),
		\\
		\M[(066)]{M}{s,Regge} &= \frac{\alpha^2}{\alpha_0}\left(\frac{1}{4}\sqrt{\frac{3}{2}}	-\frac{\alpha }{54}	-\frac{751 \alpha ^2}{10584 \sqrt{6}}	-\frac{101 \alpha ^3}{535815} + \frac{9044053 \alpha ^4}{1228730958 \sqrt{6}}\right),
		\\
		\M[(077)]{M}{s,Regge} &= \frac{\alpha^2}{\alpha_0}\left(\frac{3}{14}\sqrt{\frac{3}{2}}	-\frac{2 \alpha }{147}	-\frac{7865 \alpha ^2}{175616 \sqrt{6}}	-\frac{179 \alpha ^3}{2074464}	+ \frac{1534182161 \alpha ^4}{446092738560 \sqrt{6}}\right).
	\end{align}
\end{subequations}
%%%%%

%%%
\vskip 0.5cm
\noindent
{\bf (9) Approximate formulae for the mass corrections of dominant modes}
\\
%%%%%=====
\begin{subequations}\label{eq:delta-Ms_Regge_0ll}
	\begin{align}
		\delta\M[(011)]{M}{s} &= -\frac{3 \left(484+9 \pi ^2\right) \alpha _2^{16} \text{$\mu$ }\tau _{\text{acc}}}{327680 \alpha _0} \approx -5.24\times10^{-3} \frac{\mu\M[]\tau{acc}}{\alpha_0} \alpha^{16},
		\\
		\delta\M[(022)]{M}{s} &= -\frac{\left(1024+49 \pi ^2\right) \alpha _2^{20} \text{$\mu$ }\tau _{\text{acc}}}{128566206720 \alpha _0} \approx -1.17\times10^{-8} \frac{\mu\M[]\tau{acc}}{\alpha_0} \alpha^{20},
		\\
		\delta\M[(033)]{M}{s} &= -\frac{11 \alpha _2^{24} \text{$\mu$ }\tau _{\text{acc}}}{246230475079680 \alpha _0} \approx -4.47\times10^{-14} \frac{\mu\M[]\tau{acc}}{\alpha_0} \alpha^{24},
		\\
		\delta\M[(044)]{M}{s} &=-\frac{\left(16777216+2029052025 \pi ^2\right) \alpha _2^{28} \text{$\mu$ }\tau _{\text{acc}}}{2791851562500000000000000000 \alpha _0} \approx -7.18\times10^{-18} \frac{\mu\M[]\tau{acc}}{\alpha_0} \alpha^{28},
		\\
		\delta\M[(055)]{M}{s} &= -\frac{7 \left(210453397504+480020337225 \pi ^2\right) \alpha _2^{32} \text{$\mu$ }\tau _{\text{acc}}}{34079635094274801587650648080384000 \alpha _0} \approx -1.02\times10^{-21} \frac{\mu\M[]\tau{acc}}{\alpha_0} \alpha^{32},
		\\
		\begin{split}
			\delta\M[(066)]{M}{s} &= -\frac{\left(53876069761024+91415073021129 \pi ^2\right) \alpha _2^{36} \text{$\mu$ }\tau _{\text{acc}}}{16320026746151167744397770322620658483200 \alpha _0} 
			\\
			&\approx -5.86\times10^{-26} \frac{\mu\M[]\tau{acc}}{\alpha_0} \alpha^{36},
		\end{split}
		\\
		\begin{split}
			\delta\M[(077)]{M}{s} &= -\frac{6171 \left(17592186044416+28821275417025 \pi ^2\right) \alpha _2^{40} \text{$\mu$ }\tau _{\text{acc}}}{961449231303759399228736679876164935336171929600 \alpha _0} 
			\\
			&\approx -1.94\times10^{-30} \frac{\mu\M[]\tau{acc}}{\alpha_0} \alpha^{40}.
		\end{split}
	\end{align}
\end{subequations}
%%%%%

%%%
\vskip 0.5cm
\noindent
{\bf (10) Approximate formulae for the mass couplings when the BH mergers onto the $l>l_c$ Regge trajectory from the the Regge trajectory of the previous stage}
\\
%%%%%=====
\begin{subequations}\label{eq:alpha_1_0ll}
	\begin{align}
		\alpha_1^{(022)} &\approx \left\{-\frac{5\cdot3^{12}}{2^{7}\tau_\mathrm{acc}\mu}   W_{-1}\left[-\frac{{2}^{7}\tau_{\mathrm{acc}}\mu}{{3}^{12}\cdot{5}}\left(M_{\mathrm{s0}}^{(022)}\mu e^{-\frac{{2}^{7}\left(\alpha_{1}^{(011)}\right)^{13}\tau_{\mathrm{acc}}\mu}{{3}^{11}\cdot{5}\cdot{13}}}\right)^{{13}/{3}}\right]\right\}^{{1}/{13}},
		\\
		\alpha_1^{(033)} &\approx \left\{-\frac{{2}^{13}\cdot{3}^{3}\cdot{5}^{3}\cdot{7}}{\tau_\mathrm{acc}\mu} W_{-1}\left[-\frac{{3}^{25/3}\tau_{\mathrm{acc}}\mu}{{2}^{56/3}\cdot{5}^{3}\cdot{7}}\left(M_{\mathrm{s0}}^{(033)}\mu e^{-\frac{\left(\alpha_{1}^{(022)}\right)^{17}\tau_{\mathrm{acc}}\mu}{{2}^{13}\cdot{3}^{2}\cdot{5}^{3}\cdot{7}\cdot{17}}}\right)^{{17}/{3}}\right]\right\}^{{1}/{17}},
		\\
		\alpha_1^{(044)} &\approx \left\{-\frac{{3}^{6}\cdot{5}^{15}\cdot{7}^{3}}{{2}^{11}\tau_\mathrm{acc}\mu} W_{-1}\left[-\frac{{2}^{25}\cdot{3}\tau_{\mathrm{acc}}\mu}{{5}^{15}\cdot{7}^{3}}\left(M_{\mathrm{s0}}^{(044)}\mu e^{-\frac{{2}^{11}\left(\alpha_{1}^{(033)}\right)^{21}\tau_{\mathrm{acc}}\mu}{{3}^{6}\cdot{5}^{15}\cdot{7}^{4}}}\right)^{7}\right]\right\}^{{1}/{21}},
		\\
		\begin{split}
			\alpha_1^{(055)} &\approx \left(
			\frac{{2}^{4}\cdot{3}^{23}\cdot{5}^{2}\cdot{7}^{3}\cdot{11}}{\tau_\mathrm{acc}\mu}
			\right) ^{{1}/{25}}
			\\
			& \hspace{0.5cm} \times \left\{-W_{-1}\left[-\frac{{5}^{44/3}\tau_{\mathrm{acc}}\mu}{{2}^{4}\cdot{3}^{23}\cdot{7}^{3}\cdot{11}}\left(M_{\mathrm{s0}}^{(055)}\mu e^{-\frac{\left(\alpha_{1}^{(044)}\right)^{25}\tau_{\mathrm{acc}}\mu}{{2}^{4}\cdot{3}^{22}\cdot{5}^{4}\cdot{7}^{3}\cdot{11}}}\right)^{{25}/{3}}\right]\right\}^{{1}/{25}},
		\end{split}
		\\
		\begin{split}
			\alpha_1^{(066)} &\approx \left(\frac{{3}^{8}\cdot{5}^{3}\cdot{7}^{19}\cdot{11}^{3}\cdot{13}}{{2}^{14}\tau_\mathrm{acc}\mu}\right)^{{1}/{29}}
			\\
			& \hspace{0.5cm} \times  \left\{ - W_{-1}\left[-\frac{{2}^{14}\cdot{3}^{34/3}\cdot{5}^{20/3}\tau_{\mathrm{acc}}\mu}{{7}^{19}\cdot{11}^{3}\cdot{13}}\left(M_{\mathrm{s0}}^{(066)}\mu e^{-\frac{{2}^{14}\left(\alpha_{1}^{(055)}\right)^{29}\tau_{\mathrm{acc}}\mu}{{3}^{7}\cdot{5}^{3}\cdot{7}^{19}\cdot{11}^{3}\cdot{13}\cdot{29}}}\right)^{{29}/{3}}\right]\right\}^{{1}/{29}},
		\end{split}
		\\
		\begin{split}
			\alpha_1^{(077)} &\approx \left(\frac{{2}^{40}\cdot{3}^{10}\cdot{5}^{3}\cdot{7}^{2}\cdot{11}^{3}\cdot{13}^{3}}{\tau_\mathrm{acc}\mu}\right)^{1/33}
			\\
			& \hspace{0.5cm} \times \left\{-W_{-1}\left[-\frac{{3}\cdot{7}^{20}\tau_{\mathrm{acc}}\mu}{{2}^{51}\cdot{5}^{3}\cdot{11}^{3}\cdot{13}^{3}}\left(M_{\mathrm{s0}}^{(077)}\mu e^{-\frac{\left(\alpha_{1}^{(066)}\right)^{33}\tau_{\mathrm{acc}}\mu}{{2}^{40}\cdot{3}^{10}\cdot{5}^{3}\cdot{7}^{2}\cdot{11}^{3}\cdot{13}^{3}}}\right)^{{33}/{3}}\right]\right\}^{{1}/{33}}.	
		\end{split}
	\end{align}
\end{subequations}
%%%%%

%%%
\vskip 0.5cm
\noindent
{\bf (11) Approximate formulae for the mass couplings when $\M[(0ll)]{M}{s}/M$ reaches its maximum}
\\
%%%%%=====
\begin{subequations}\label{eq:alpha_2_0ll}
	\begin{align}
		\alpha_2^{(011)} &= \left(\frac{5120 \sqrt{6}}{9 \pi ^2 \mu \tau _{\text{acc}}+484 \mu \tau _{\text{acc}}}\right)^{1/14} \approx 1.24\times\left(\mu \tau _{\text{acc}}\right)^{-1/14}\\
		\alpha_2^{(022)} &= \left(\frac{2410616376 \sqrt{6}}{49 \pi ^2 \mu \tau _{\text{acc}}+1024 \mu \tau _{\text{acc}}}\right)^{1/18} \approx 2.32\times\left(\mu \tau _{\text{acc}}\right)^{-1/18},\\
		\alpha_2^{(033)} &= \left(\frac{2564900782080 \sqrt{6}}{11 \mu \tau _{\text{acc}}}\right)^{1/22} \approx 3.42\times\left(\mu \tau _{\text{acc}}\right)^{-1/22},\\
		\alpha_2^{(044)} &= \left(\frac{18695434570312500000000000 \sqrt{6}}{2029052025 \pi ^2 \mu \tau _{\text{acc}}+16777216 \mu \tau _{\text{acc}}}\right)^{1/26}\approx 3.90\times\left(\mu \tau _{\text{acc}}\right)^{-1/26},\\
		\alpha_2^{(055)} &= \left(\frac{159748289504413132442112412876800 \sqrt{6}}{3360142360575 \pi ^2 \mu \tau _{\text{acc}}+1473173782528 \mu \tau _{\text{acc}}}\right)^{1/30}\approx 4.32\times\left(\mu \tau _{\text{acc}}\right)^{-1/30},\\
		\begin{split}
			\alpha_2^{(066)} &= \left(\frac{510000835817223992012430322581895577600 \sqrt{\frac{2}{3}}}{274245219063387 \pi ^2 \mu \tau _{\text{acc}}+161628209283072 \mu \tau _{\text{acc}}}\right)^{1/34} 
			\\
			&\approx 4.80\times\left(\mu \tau _{\text{acc}}\right)^{-1/34},
		\end{split}
		\\
		\begin{split}
			\alpha_2^{(077)} &= \left(\frac{858436813664070892168514892746575835121582080 \sqrt{6}}{59285363532820425 \pi ^2 \mu \tau_{\text{acc}}+36187126693363712 \mu \tau_{\text{acc}}}\right)^{1/38} 
			\\
			&\approx 5.30\times\left(\mu \tau _{\text{acc}}\right)^{-1/38}.
		\end{split}
	\end{align}
\end{subequations}
%%%%%

%%%
\vskip 0.5cm
\noindent
{\bf (12) Approximate formulae for the mass couplings when $\M[(1ll)]{M}{s}$ reaches its maximum}
\\
%%%%%=====
\begin{subequations}\label{eq:alpha_2_1ll}
	\begin{align}
		\alpha_2^{(111)} &= \left(\frac{864}{5 \mu \tau _{\text{acc}}}\right)^{1/11} \approx 1.60\times\left(\mu \tau _{\text{acc}}\right)^{-1/11}\\
		\alpha_2^{(022)} &= \left(\frac{7971615}{14 \mu \tau _{\text{acc}}}\right)^{1/15} \approx 2.42\times\left(\mu \tau _{\text{acc}}\right)^{-1/15},\\
		\alpha_2^{(033)} &= \left(\frac{5734400000}{\mu \tau _{\text{acc}}}\right)^{1/19} \approx 3.26\times\left(\mu \tau _{\text{acc}}\right)^{-1/19},\\
		\alpha_2^{(044)} &= \left(\frac{190770721435546875}{1408 \mu \tau _{\text{acc}}}\right)^{1/23}\approx 4.12\times\left(\mu \tau _{\text{acc}}\right)^{-1/23},\\
		\alpha_2^{(055)} &= \left(\frac{83543560665596539200}{13 \mu \tau _{\text{acc}}}\right)^{1/27}\approx 4.97\times\left(\mu \tau _{\text{acc}}\right)^{-1/27},\\
		\alpha_2^{(066)} &= \left(\frac{35227172069319515266026045}{64 \mu \tau _{\text{acc}}}\right)^{1/31} \approx 5.83\times\left(\mu \tau _{\text{acc}}\right)^{-1/31},\\
		\alpha_2^{(077)} &= \left(\frac{1339612089633916391635550208000}{17 \mu \tau _{\text{acc}}}\right)^{1/35} \approx 6.69\times\left(\mu \tau _{\text{acc}}\right)^{-1/35}.\\
	\end{align}
\end{subequations}
%%%%%

%%%%%%%%%%%%%%%%%%%%%%%%%%%%%%%%%%%%%%%%%%
%%%%%%%%%% Bibliography %%%%%%%%%%%%%%%%%%
%%%%%%%%%%%%%%%%%%%%%%%%%%%%%%%%%%%%%%%%%%

%\bibliography{reference}

\end{document}